\documentclass[
nofootinbib,
 amsmath,amssymb,
 aps,twocolumn
]{revtex4}
\usepackage[parfill]{parskip}
\usepackage{graphicx}
\usepackage{amssymb}
\usepackage{amsmath}
\usepackage{mathtools}
\usepackage{tikz}
\usepackage{color}
\usepackage{caption}
\usepackage{hyperref}
\usepackage{ulem}

\newcommand{\be}{\begin{equation}}
\newcommand{\ee}{\end{equation}}
\newcommand{\ba}{\begin{eqnarray}}
\newcommand{\ea}{\end{eqnarray}}

\begin{document}
\title{Ridges in rotating neutron--star properties due to first order phase transitions }
\author{Pablo Navarro Moreno, Felipe J. Llanes--Estrada and Eva Lope--Oter}
\affiliation{Univ. Complutense de Madrid, dept. F\'isica Te\'orica and IPARCOS, Plaza de las Ciencias 1, 28040 Madrid, Spain.}

\begin{abstract}

We identify combinations of observables  for rotating neutron stars that can one day bear on the question of 
whether there can be first order phase transitions in the neutron matter therein. 
We employ the Hartle-Thorne theory for stationary, rotating neutron stars at conventional angular 
velocities (in the pulsar and millisecond pulsar ranges) and extract three-dimensional sections of
the ellipticity or the dynamical angular momentum as function of the star's mass and angular velocity.
An eventual first order phase transition in the equation of state (EoS) leaves a clear ridge (nonanalyticity)
in these observables, akin to the sudden kink in popular mass-radius diagrams for static stars.
Finally, we observe that static neutron stars in General Relativity (GR) will fail to be compact enough 
for the light ring's position at $r=3M$ to be outside the star, except for the most extreme equations of state. 
The outer light ring of a rotating star might however be formed unless the EoS softens too much, and its eventual detection can then  
be used to constrain the EoS (or the gravity theory).

\end{abstract}

\maketitle

\section{Introduction: searching for phase transitions in neutron star matter with rotating stars}

The equation of state of neutron star matter has become a central problem of nuclear and particle physics~\cite{Llanes-Estrada:2019wmz,Burgio:2021vgk} and systematic efforts have started aimed at ordering all available information~\cite{MUSES:2023hyz}
for optimal use of astrophysical facilities.

One of the more interesting questions about it is whether a phase transition to a nonhadronic \cite{Alford:1997zt,GomezDumm:2006xf,Gorda:2021znl,AngelesPerez-Garcia:2022qzs} (or at least,
with significant strangeness~\cite{Vidana:2000ew,Blacker:2023opp}) phase can be present in physical neutron stars and not only on theoretical models at unreachable densities.

The discovery of very massive neutron stars above the two--solar mass line is making the case for such exotic
phases more difficult. But a modest additional support against gravitational collapse can be provided by rotation. 
The interest in rotating stars has also picked up~\cite{Posada:2023bnm} as gravitational wave detectors 
can constrain their spin in binary systems and because the residual object left after the binary collision can 
be a fast rotator~\cite{Hanauske:2021rjk}.

The literature already features observables of rotating neutron stars that can be searched for and eventually 
impact the search for deconfined quark matter or other exotic phases of hadron physics, such as the proposed backbending 
phenomenon due to a change in the moment of inertia~\cite{Glendenning:1997fy,Bejger:2016emu,Franzon:2016urz}  
that can lead to spin up of the star.

Detailed fits to large families of stars will be painstaking and also can get confused by the degeneracy with modified
gravity, that can present similar effects to exotic hadron phases. 
It is thus also of interest to design observables that are smoking guns of phase transitions. Because first order phase 
transitions present a latent heat and therefore force nonanalyticities (kinks) in observable-to-observable diagrams, such as 
the mass-radius diagram (care has to be taken with masquerading stars~\cite{Wei:2018mxy} and the possibility of scalarization phase transitions within the theory of gravity~\cite{Doneva:2023kkz}), we dedicate this article
to study said nonanalyticities in the accessible observables of rotating neutron stars.

Although we have tried to have a self-contained article that can be read by a nuclear physicist and by an astrophysicist,
much detail must be left out in the interest of conciseness, so we refer to recent reviews on rotating stars, the equation of state and gravitational wave detection~\cite{Paschalidis:2016vmz,Llanes-Estrada:2019wmz,Burgio:2021vgk,Cahillane:2022pqm}

We focus on observables related to rotating neutron stars. The spin period of several pulsars, measured from the interval between
the successive pulses, are presented in table~\ref{Expperiods}. 
\begin{table}[h!]
\centering
\begin{tabular}{|c|c|c|c|}
\hline
Pulsar & $F_0(Hz)$ & $\Omega({\rm rad}\cdot {\rm ms}^{-1})$ & $M(M_{\odot})$ \\ \hline
J0337+1715 & 365.953 & 2.299 & 1.4401(15) \\ \hline
J1012+5307 & 190.267 & 1.195 & 1.72(16) \\ \hline
J0348+0432 & 25.561 & 0.161 & 2.01(4) \\ \hline
J0453+1559 & 21.843 & 0.137 & 1.559(5) \\ \hline
J0509+380  & 13.065 & 0.082 & 1.34(8) \\ \hline
\end{tabular}
\caption{Frequency, angular velocity and mass of a few well-measured pulsars~\cite{Antoniadis:2016hxz}. \label{Expperiods}}
\end{table}

Observed neutron star masses lie in the rather compact interval 1-2.3$M_{\odot}$,
although the nEoS band of equations of state that we deploy, described in the appendix, allows for neutron star masses to extend 
upwards of three solar masses in General Relativity. This is consistent with the information from microscopic physics alone, having
employed no astrophysical constraint, so our plots will extend beyond the GR range (also keep in mind that gravity might need to be tested and perhaps modified in the intensely dense neutron--star environment).
There is more spread in the observed period of pulsars, that range from millisecond to seconds. The fastest known pulsar has a frequency of 716 Hz that corresponds to  $\Omega=4.5\;\rm{ms^{-1}}$. We will therefore illustrate our computations with angular velocities up to 
$\Omega= 5$ radians per millisecond.

In General Relativity, the maximum angular velocity above which the centrifugal force exceeds the gravitational pull, so that the star sheds mass, is~\cite{Haensel:2009wa,Glendenning} 
\begin{equation}\label{Omegak}
    \Omega_k\approx 0.65\sqrt{\frac{M}{R^3}}.
\end{equation}
That is reminiscent of the Newtonian mechanics expression $ \Omega_c=\sqrt{\frac{M}{R^3}}$ obtained by the balance of forces
$\frac{Mm}{R^2}=m\Omega_c^2R$.
Equation~\ref{Omegak} can be rewritten as
\begin{equation}
     \Omega_k\approx 240\left(\frac{M/M_{\odot}}{(R/\rm{km})^3}\right)^{1/2}\rm{ms^{-1}}
\end{equation}
and yields a bound on angular velocity above which it stops making sense to plot observable quantities, as no
stationary star will populate such high $\Omega$ values.

\begin{figure}[h!]
\includegraphics[width=0.8\linewidth]{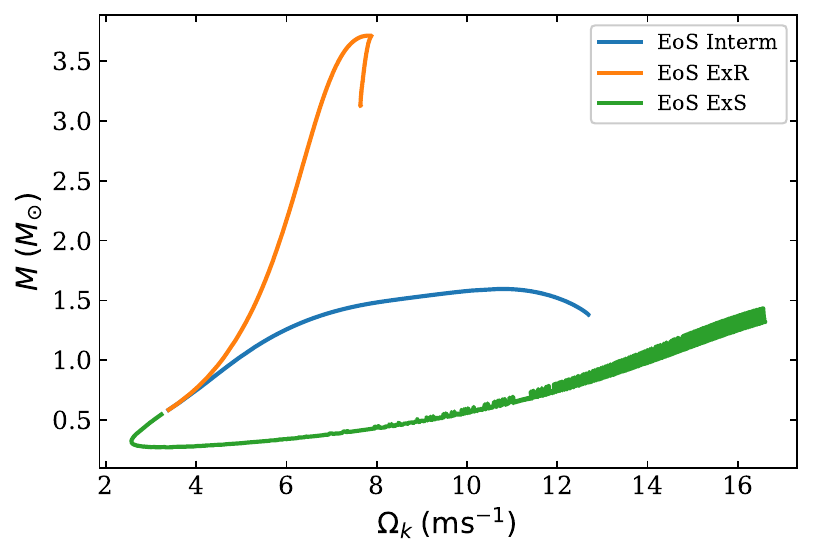}
\caption{Maximum Keplerian velocity at mass shedding against the total star mass for three equations of state, as obtained from Eq.~(\ref{Omegak}).
}
\label{fig:Keplerw}
\end{figure}

Figure~\ref{fig:Keplerw} plots that maximum Keplerian velocity versus the neutron star mass for three computed families of stars 
with different equations of state that are extremely hard (ExR), extremely soft (ExS) and intermediate in the sense 
explained below around figure~\ref{eosfig}.

None of our computations will of course exceed that Keplerian velocity.  But moreover, we will adopt the Hartle-Thorne slow
rotation condition which does not cover the Kepler limit leading to the mass-shedding region.
To be specific, we would like to maintain $\Omega R<0.05$ which includes most pulsars (see the discussion below in section~\ref{sec:HartleThorne}).

Throughout this work we employ the geometrodynamics system $G=1=c$. Then the conversion from solar masses to kilometers is
$1M_\odot = 1.47$ km, and the connection to the microscopic natural units (with different dimensionality, $E\sim L^{-1}$ instead of
$E\sim L$ as in the geometrized one), which is
necessary to employ conventional equations of state,
is obtained by the substitution
\begin{equation}\label{Unidades7}
    \frac{\rm{MeV}}{\rm{fm}^3}\to 1,31752\cdot 10^{-6}\frac{1}{\rm{km}^2}\ .
\end{equation}

A brief revision on static stars to settle notation in section~\ref{sec:static} is followed
by the Hartle-Thorne theory of rotating stars, in section~\ref{sec:HartleThorne}. the computation of basic 
observable quantities in section~\ref{sec:extraction} and the demonstration of nonanalyticities in them (section~\ref{sec:numerics})
are the heart of the work.
Section~\ref{sec:lightring} then discusses what the chances of detecting a light ring are, and the discussion closes with an 
outlook in section~\ref{sec:outlook}.

\section{Static stars from hadron physics alone}\label{sec:static}
A static, spherically symmetric star can be described by the inner Schwarzschild metric 
\begin{equation}\label{metrica}
    ds^2=-e^{\nu(r)}dt^2+e^{\lambda(r)}dr^2+r^2(d\theta^2+\sin^2{\theta}d\phi^2),
\end{equation}
with $e^{\nu(r)}=1-\frac{2M}{r}= e^{-\lambda(r)}$. Einstein's equations with an ideal fluid
characterized by $(\rho,P)$ (energy density, also called $\varepsilon$, and pressure, respectively) and a 
barotropic $P(\rho)$ equation of state for zero temperature matter lead to the Tolman-Oppenheimer-Volkoff system
\begin{eqnarray}\label{TOV1}
    \frac{dm(r)}{dr}=4\pi r^2\rho(r),\\
\label{TOV2}
    -\frac{dp(r)}{dr}=(\rho(r)+p(r))\frac{4\pi r^3p(r)+m(r)}{r^2(1-2m(r)/r)},
\\
\label{TOV3}
    \frac{1}{2}\frac{d\nu(r)}{dr}=-\frac{1}{\rho(r)+p(r)}\frac{dp(r)}{dr}.
\end{eqnarray}
That provides the metric function $\nu(r)$, the quantity of energy-matter $m(r)$ and the pressure $P(r)$ as function of the distance to the star's center once integrated by Runge-Kutta. The total star mass as read off by the external Schwarzschild metric coincides with 
the matter accrued $M=m(R)$ up to the surface characterized by $P=0$.

\begin{figure}
\centering
\includegraphics[width=0.9\columnwidth]{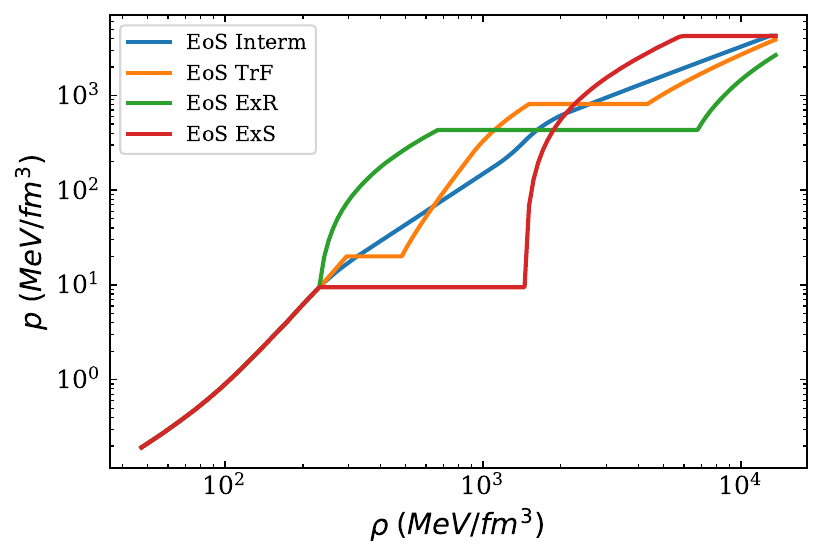}
\caption{The nEoS band of barotropic equations of state $P(\rho)$ in log scale, which is a sophisticated version of the traditional
Rhoades-Ruffini~\cite{Rhoades:1974fn} diagram taking into account pQCD information~\cite{Kurkela:2022elj}. 
See appendix~\ref{app:eos} for additional discussion.
 }
\label{eosfig}
\end{figure}

It remains to specify the EoS $P(\rho)$. Figure~(\ref{eosfig}) shows the nEoS\footnote{\tt http://teorica.fis.ucm.es/nEoS } band within which the actual equation of state of neutron stars has to lie irregardless of what the correct theory of gravity is, as the plot includes information only from hadron physics.
The uncertainty on the lowest density interval is not visible on this scale because data from nuclear laboratory observables,
extrapolations based on direct calculations with the chiral Lagrangian, and realistic nuclear potentials constrain 
the EoS. At the top right corner of the figure we have the perturbative Quantum Chromodynamics (QCD) constrains. 

The visible uncertainty band is actually a twisted tube in a four-dimensional space ($\rho$, $P$, $\mu$, $n$) that includes the chemical potential and number density, and
satisfying an integral thermodynamic constraint $P=\int n d\mu$. This complicates this two-dimensional projection, that would otherwise
have as boundaries four curved segments given by the extreme sound speeds allowed by causality $c_s^2 = dP/d\rho \leq 1$ and
stability $c_s^2=dP/d\rho \geq 0$, as in the Rhoades-Ruffini diagram.

A stretch of zero derivative in this plot represents a first order phase transition with a latent heat: an infinitesimal
increase in the pressure $P$ entails a finite jump in the energy density $\rho$. This nonanalyticity is precisely the feature
that we would like to extract from astrophysical observables, where it would be reflected as a jump in stellar properties. 
This jump is limited by hadron physics~\cite{Lope-Oter:2021mjp}, but within static stars in General Relativity, there is a tighter limit due to Seidov~\cite{Seidov}.
Our deployed mock phase transitions satisfy those limits (actually, due to particularities of the grid, the EoS with a long phase transition exceeds it by 2\% or about $10{\rm MeV}/{fm^3}$ which is small enough to be of no further concern).

\begin{figure}
\includegraphics[width=0.9\columnwidth]{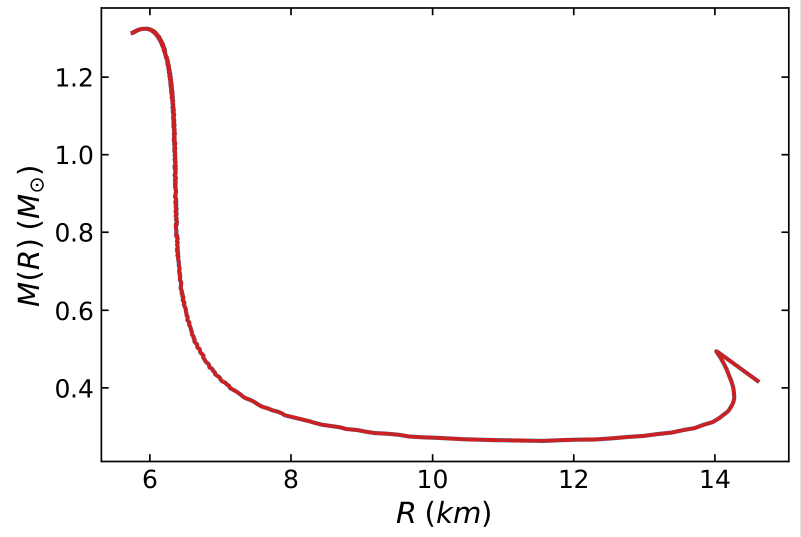}
\includegraphics[width=0.9\columnwidth]{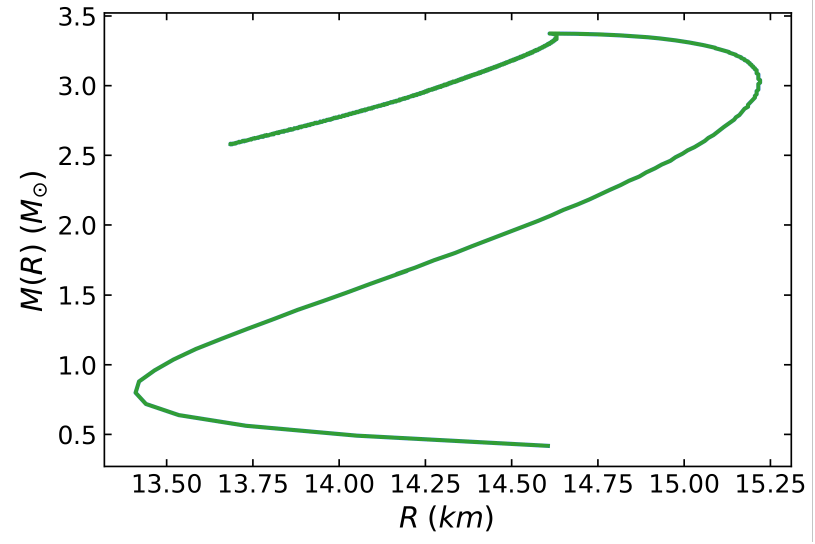}
\caption{\label{fig:static} Traditional mass-radius diagram within General Relativity for the softest and hardest EoS from 
the nEoS band allowed by hadron physics alone, without constraints from astrophysical measurements.}
\end{figure}

We use our own static Tolman-Oppenheimer-Volkoff numerical solver, which will be useful as a check of our separate rotating star
solver that  later employed in section~\ref{sec:numerics}. The extreme mass-radius lines corresponding to the green and red lines 
in the EoS figure~\ref{eosfig} are then rendered in figure~\ref{fig:static}.

It is clear that the softest nEoS equation (left plot in the figure) is too soft {\it if General Relativity is unchanged inside neutron stars}, while the hardest equation of state (right plot) yields a seemingly too high maximum mass above 3 $M_\odot$ also within General Relativity. Many groups have worked out the astrophysics constraints that dent into the EoS space of figure~\ref{eosfig}, 
but we leave out astrophysical observables in the construction of nEoS sets to guarantee their usability with modified gravity theories~\cite{Staykov:2023ose}.
The interested reader can find a recent computation of the 95\% contour allowed in this plot upon imposing astrophysical constraints within GR in~\cite{Musolino:2023edi}.

More to the point of this article, and irrespectively of the maximum value of the mass and canonical neutron star radii, 
we can see that the phase transition (flat stretch of the  red line  in figure~\ref{eosfig}, the bottom one at 400 MeV/fm$^3$) causes a clear kink in the bottom right corner of the top plot of figure~\ref{fig:static}. That is, the $M(R)$ diagram (and also $M(I)$ for a rotating star as defined shortly) shows phase transitions of the equation of state as kinks~\cite{Paschalidis:2017qmb,Bauswein:2018bma,Han:2018mtj,Most:2018eaw}. 
Likewise, the flat stretch of the top green line in figure~\ref{eosfig} yields another kink in the bottom plot of figure~~\ref{fig:static}
at about the top of the curve, where a sudden change in derivative is clearly visible.
Similar nonanalyticities will appear in the observables of rotating stars as we will here show.

\section{Stationary rotating stars} \label{sec:HartleThorne}

This static star is then modified by rotation following the theory of Hartle and Thorne~\cite{Hartle:1967he,Hartle:1968si} that we briefly expose in appendix~\ref{app:rotation}. 
The metric with a stationary time dependence is
\begin{equation}\label{metrica2}
    ds^2=-H^2dt^2+Q^2dr^2+r^2K^2(d\theta^2+\sin^2{\theta}(d\phi-Ldt)^2).
\end{equation}
where $H,Q,K,L$ are functions of the variables $(r,\theta)$.

The angular velocity of a star is contingent on many variables related to its birth, accretion if any and radiation, so for a given star mass there is a family of neutron stars tagged by an appropriate rotation parameter $\Omega$. 
This is conventionally taken as the angular velocity of the star's surface as seen from an observer at infinity. 
It differs from the local angular velocity $L$ at each point of the star, that appears in the metric of Eq.~(\ref{metrica2}) and controls
(through the Christoffel symbols) the inertial (centrifugal, Coriolis, and drag) accelerations. There is no azimuthal acceleration as the rotation is considered stationary on observational time scales. 

All pulsars that satisfy this condition (that is, all known pulsars except at brief instants called ``glitches'' and during binary collisions) also happen to be slow rotators in the sense of Hartle and Thorne, that is, 
the angular velocity parameter does not displace any point of the star at velocities near $c$, 
\begin{equation}\label{slowrot}
    \Omega R \ll c\ .
\end{equation}
Then, the effect of the rotation can be perturbatively added to the quantities that solve the TOV static system. 
No pulsar currently known appears to rotate with frequency exceeding a kilohertz; if we take as an  example of a quickly rotating object 
the millisecond pulsar PSR J0030+0451 with a frequency of $f=205.53$ Hz~\cite{Riley:2019yda,Raaijmakers:2019qny}],
the surface velocity as seen from infinity is still within the reasonable applicability range of the approximation with $\frac{\Omega R}{c}=0.055$. A Taylor expansion in powers of $\Omega$ of all quantities is then warranted.
The fastest rotator, at 716 Hz, would in turn have $\frac{\Omega R}{c}$ of order $15$ to $25\%$ depending on the radius
and further theoretical refinements could be necessary. 

The star is distorted, with a breathing or radial mode $\xi_0$ and a quadrupolar mode $\xi_2$ (see figure~\ref{deformations})
that leads to an ellipticity of the star's shape, which we denote as $e$. 
The numerical extraction, based on the lean theoretical material given in appendix~\ref{app:rotation} is presented in the next
section~\ref{sec:extraction}.

\section{Theoretical extraction of observable quantities} \label{sec:extraction}
\subsection{Ellipticity}
After the numeric integration of the constitutive equations of a rotating star in appendix~\ref{app:rotation},  $\xi_0(r)$ and $\xi_2(r)$ are reconstructed from the auxiliary quantities $p_0^*$ and $p_2^*$,
and we have then at hand the polar and equatorial radii of the rotationally deformed star,
by respectively evaluating Eq.~(\ref{rp}) at $\theta=0$ and $\theta=\pi/2$. 

This leaves
\begin{eqnarray}\label{rpolar}
    r_{\rm{polar}}=R+\xi_0(R)+\xi_2(R),
\\ \label{req}
    r_{\rm{eq}}=R+\xi_0(R)-\frac{1}{2}\xi_2(R)
\end{eqnarray}
from which the star's ellipticity can be computed
\begin{equation}\label{e}
    e=\sqrt{1-\left(\frac{r_{\rm{polar}}}{r_{\rm{eq}}}\right)^2} .
\end{equation}

Currently we are not aware of any measurement of a nonvanishing ellipticity for a pulsar, 
but the radius of a neutron star, with a network of three next-generation gravitational wave detectors such as 
the Einstein telescope, will reach precisions down to the 50 meter level~\cite{Huxford:2023qne,Branchesi:2023mws}, and then ellipticities will also be  accessible, for example combining studies of the stochastic gravitational-wave background~\cite{Talukder:2014eba}.
It should be observed that the quadrupole moment of an isolated star  does not change if the spin and quadrupole's $OZ$ axis are aligned. Still, in a binary system, if the spins
      of the two objects are not aligned with the orbit, the quadrupole will receive a contribution from the ellipticity of either body.
      We ignore whether such misaligned neutron star pairs have been studied in detail, but there is extant work for black hole pairs~\cite{Fishbach:2022lzq,Cattorini:2022tvx} and the detection of a pair of misaligned neutron stars could help extract the 
      individual ellipticities.

\subsection{Angular momentum}
To pick one star from the family of possible solutions corresponding to a given 
equation of state, once Eq.~(\ref{eq.alpha}) and~(\ref{eq.deralpha}) have been integrated, a reference angular velocity 
at the star's surface  $\Omega$ is chosen. Then we can extract the angular momentum $J$ as follows.

In the exterior metric, the auxiliary function $j$ becomes trivial
\begin{equation}\label{j2}
    j(r)=1\;\;\;\;\;\;\;\;\;\;\forall r>R \ .
\end{equation}
Eq.~(\ref{eq.varpi3}) reduces, also outside the star, to
\begin{equation}\label{eq.varpi4}
    \frac{d}{dr}\left(r^4\frac{d\varpi}{dr}\right)=0\Longrightarrow r^4\frac{d\varpi}{dr}=\rm{constant}.
\end{equation}

From Eq.~~(\ref{eq.varpi4}) one gets, upon choosing the constant to match with the Newtonian angular momentum
\begin{equation}\label{J}
    6J=r^4\left(\frac{d\varpi}{dr}\right)\Longrightarrow J=\frac{R^4}{6}\left(\frac{d\varpi}{dr}\right)_R\ .
\end{equation}

That factor 6 is obtained by integrating Eq.~(\ref{eq.varpi3}) outwards, substituting $\displaystyle\frac{dj}{dr}$ from Eq.~(\ref{derj}) to 
yield
\begin{eqnarray}\label{J2}
    {\rm constant}=kJ&=&\left(r^4\frac{d\varpi}{dr}\right)_R
    \\ \nonumber &=&16\pi\int_0^R dr r^4\frac{(\rho+p)e^{-\nu/2}}{\sqrt{1-2m/r}}(\Omega-\omega(r)).
\end{eqnarray}
and fixing $k$ (that has to be 6) so that the expression
\begin{equation}\label{J3}
    J=\frac{8\pi}{3}\int_0^R dr r^4\frac{(\rho+p)e^{-\nu/2}}{\sqrt{1-2m/r}}(\Omega-\omega(r))
\end{equation}
coincides, upon taking $p\ll\rho$ and $2m/r\ll 1$, with the Newtonian limit in Eq.~(\ref{Jnew})
of the appendix
which allows to interpret $J$ as the angular momentum with the usual normalization.

Returning to Eq.~(\ref{eq.varpi4}) now that the constant has been determined, we can integrate it
to yield
\begin{equation}\label{varpi}
    \varpi(r)=-\frac{6J}{3r^3}+C
\end{equation}
that we will equate to the first order expression
\begin{equation}\label{varpi2}
    \varpi=\Omega-\omega
\end{equation}
to obtain, outside the star,
\begin{equation}\label{Omega}
    \omega(r)=\frac{6J}{3r^3}=\frac{2J}{r^3}\;\;\;\;\;\;\;\;\;\;\Omega=C,
\end{equation}
and finally,
\begin{equation}\label{Omega2}
    \Omega=\varpi(R)+\frac{2J}{R^3}.
\end{equation}
That relates the reference star size (from solving the static TOV system) $R$~\footnote{$R$ is not directly measurable, as an observational extraction could extract, for example
$(2R_{\rm Equatorial}+R_{Polar})/3= R+\xi_0$, and theory would be needed to disentangle $\xi_0$, as per Eq.~(\ref{xi0}) and~(\ref{eql03}),
but the difference between $R$ and $R+\xi_0$ is one higher order in the Hartle-Thorne expansion.
}
 and the angular velocity $\Omega$, allowing  the calculation of  a naturally defined moment of inertia
\begin{equation}\label{I}
    I=\frac{J}{\Omega} = \frac{R^3}{2} \frac{\omega(R)}{\Omega}
\end{equation}
(note the dimensionality of a moment of inertia in geometrized units, $[I]= [MR^2]\to [R^3]$).

\begin{figure*}[t]
\centering
\includegraphics[width=0.9\textwidth]{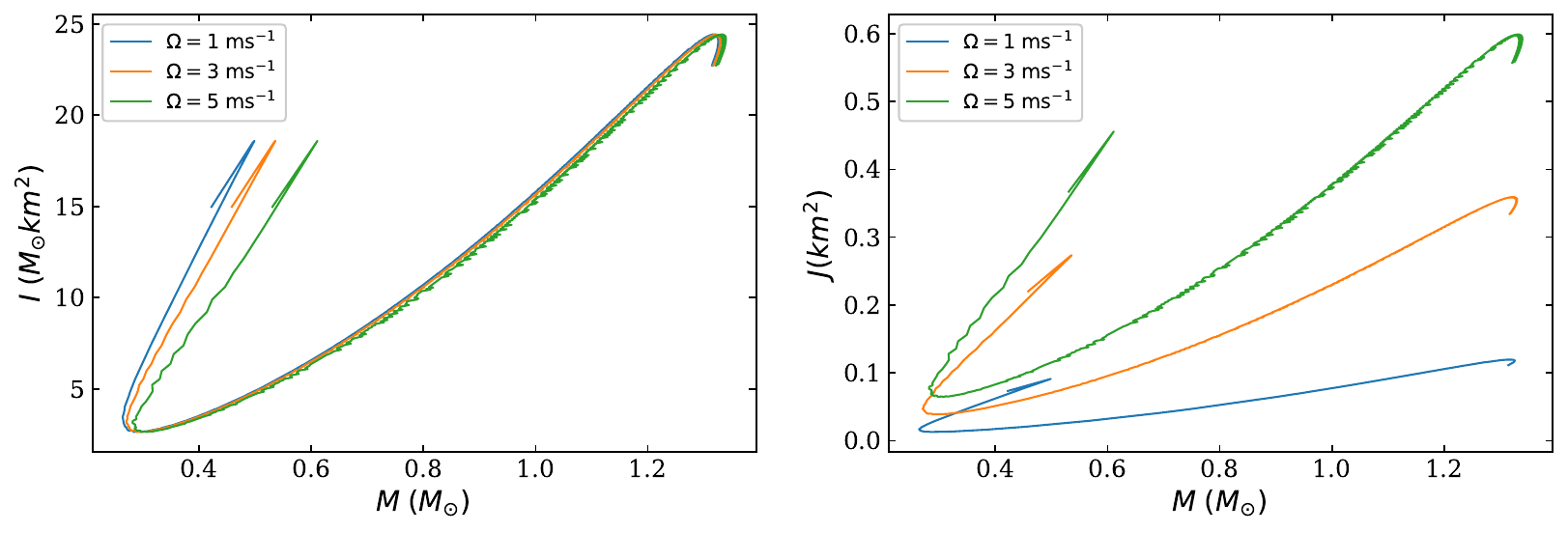}
\includegraphics[width=0.9\textwidth]{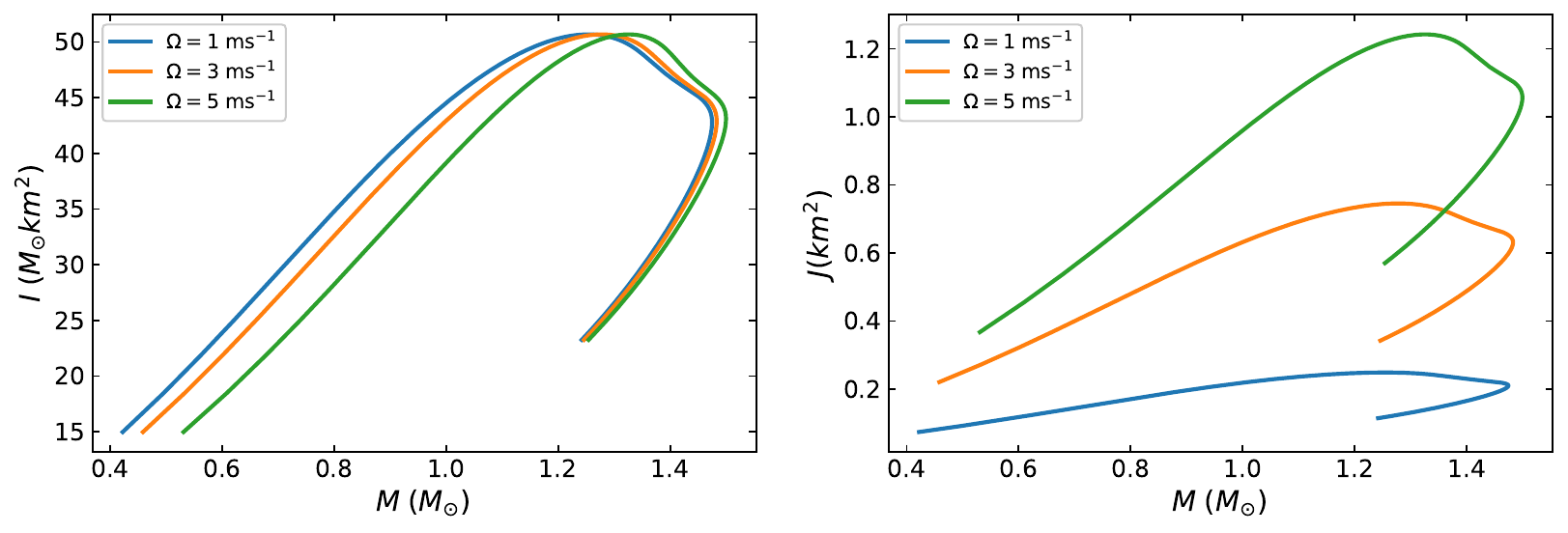}
\includegraphics[width=0.9\textwidth]{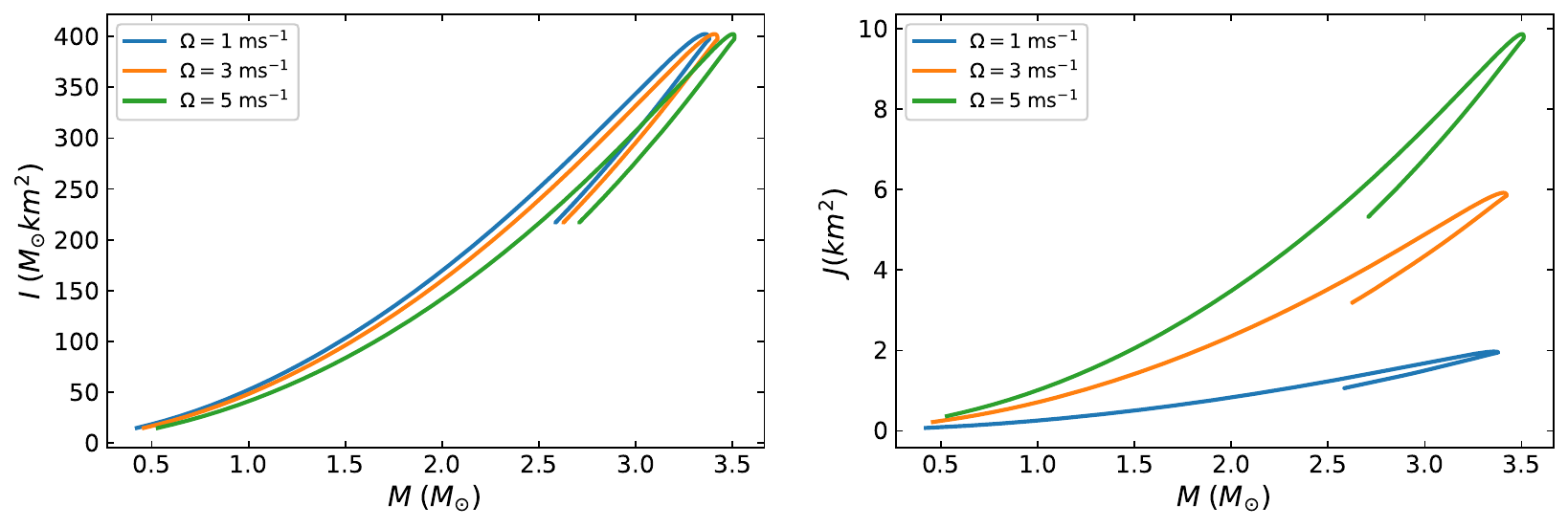}
\caption{Left column: Moment of inertia $I$. Right column: Angular momentum $J$.
Both as function of the total stellar mass $M$. From top to bottom, each row employs a different EoS from
figure~\ref{eosfig}: the softest, an intermediate one, and the hardest EoS. Three angular velocities are shown on each plot.}
\label{figIJ}
\end{figure*}

From Eq.~(\ref{J}) and (\ref{I}) we can easily obtain the angular momentum and the moment of inertia for a fixed EoS. 
We plot these in figure (\ref{figIJ}). 

From the figure, it appears that for low-mass stars, larger angular velocities entail smaller moments of inertia. This is a relativistic effect, as in classical mechanics, increasing $\Omega$ either leaves $I$ invariant or slightly increases it due to the centrifugal force that makes the star more oblate. This effect is not visible here, and the reason for that decreasing $I(\Omega)$ is the strong radial dependence
of Eq.~(\ref{I}), since the ratio $\omega/\Omega$ is universal.

As functions of the stellar mass, both  $I$ and $J$ change derivative, reflecting the behavior of the mass-radius diagram.
The total angular momentum is seen to be sensitive to the Equation of State, with the rigidmost one (marked EoS ExR in figure~(\ref{eosfig}) ) reaching angular momenta an order of magnitude larger than other EoS.

Since the $I(M)$ plots have the same shape for different equations of state, it is natural to expect universal relations~\cite{Li:2023owg} such as 
ILoveQ~\cite{Yagi:2013awa,Blazquez-Salcedo:2022pwc} that exceed the present manuscript.
In the future, theoretical knowledge of $I(M,\Omega)$ with more realistic computations could allow the study 
of accretion processes by measuring the change of angular frequency of the pulsar.

Angular momentum is a bit more directly accessible, although we cannot measure a radial free fall towards the star to measure 
its geodesic deviation dragged by the rotations. However, gravitational waves allow the inference of a combination of the two spins in a binary system, $\chi_{\rm{eff}}$~\cite{LIGOScientific:2020kqk} that is defined as
\begin{equation}\label{chi_eff}
    \chi_{\rm{eff}}=\frac{\chi_1\cos\theta_1+\chi_2\cos\theta_2}{1+q},
\end{equation}
where $q=m_2/m_1$, $\chi_i$ is the adimensional spin  defined as $\chi_i:=\left|cJ_i/(Gm_i^2)\right|$ from $J_i$, the angular momentum of object  $i$. $\chi_{\rm{eff}}$ is extractible from the gravitational wave pulse that therefore reveals an average over the
angular momentum of the two components of the binary system. For GW170817 the estimate~\cite{LIGOScientific:2018hze} for that effective spin is  $\chi_{\rm{eff}}\in (-0.01,0.17)$ employing the SEOBNRT model 
(see~\cite{Matas:2020wab} for extraction of spinning-source parameters from gravitational waves).

\section{Nonanaliticities in the presence of phase transitions} \label{sec:numerics}

\begin{figure*}[h]
\includegraphics[width=0.7\columnwidth]{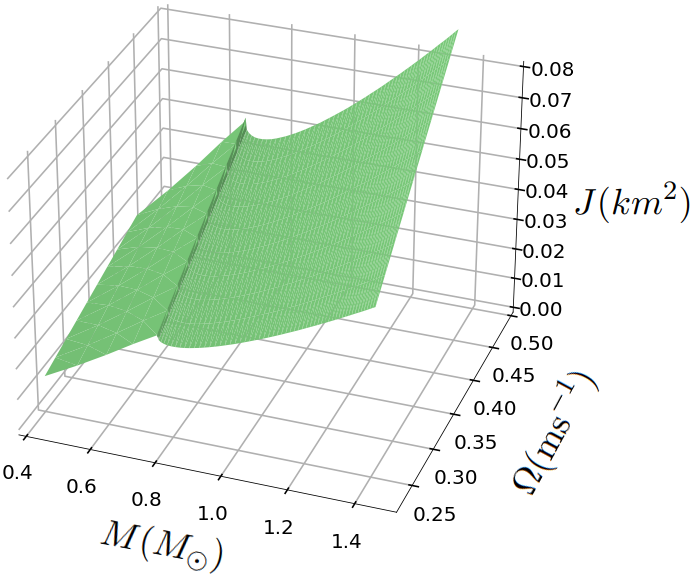}
\includegraphics[width=0.7\columnwidth]{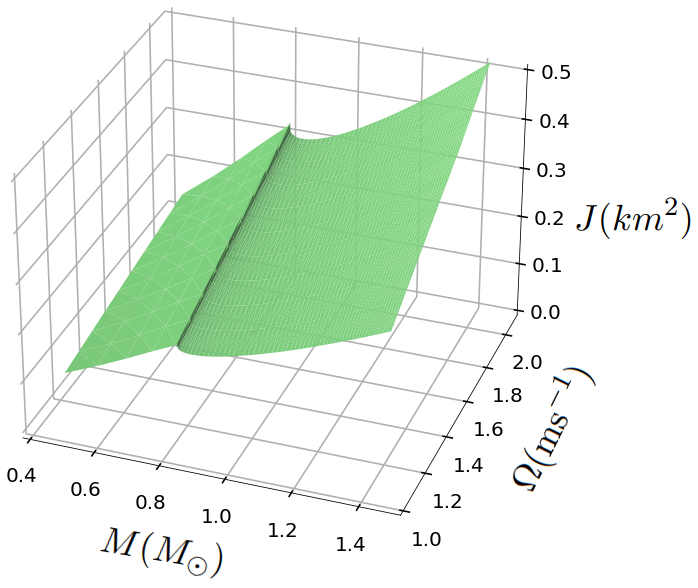}
\caption{\label{RidgeinJ} The dynamical angular momentum $J$ is predicted
to present a clear ridge (line of nonanaliticity) in the diagram plotting it against 
the neutron star mass and the reference surface angular velocity $\Omega$ (smaller 
values in the top plot, larger ones in the bottom one), should a first order 
phase transition be present in neutron star matter. The top and bottom plots offer a view
into different $\Omega$ intervals. The EoS employed is here that marked as TrF in figure~\ref{eosfig}.  }
\end{figure*}

Although it appears likely that phase transitions to phases with kaon condensates~\cite{Thapa:2020usm} or a quark core will appear, if at all~\cite{Brandes:2023hma}, only for the heaviest neutron stars,
with, for example, Mellinger {\it et al.}~\cite{Mellinger:2017dve} finding a minimum mass as function of the frequency given by 
an approximate relation $M>(1.9+0.26 \Omega^2/{\rm kHz}^2)M_\odot$, 
we are going to use a softer equation of state among those allowed from hadron physics alone  in the nEoS band~\cite{Oter:2019rqp} (with no astrophysics feedback built in). This is because in theories beyond General Relativity quark matter could appear at lower 
total masses~\cite{Lope-Oter:2023urz} given that the effective strength of gravity would be modified~\cite{Dobado:2011gd}, and for illustrational purposes, without claim of detailed predictivity. 
We set the temperature to zero as latent heat is typically largest at lower temperatures, but see {\it e.g.} \cite{Lope-Oter:2021vxl,Shaikh:2023bem} for finite temperature extensions.

We now arrive to the central result of this article, the ridges in certain plots relating observable to observable
that we wish to promote so that the observational effort can address the question of a first order phase transition in neutron star matter.

First, figure~\ref{RidgeinJ} displays a clear ridge in the plot of the total angular momentum against the star's mass and surface angular velocity, $J(M,\Omega)$.
This graph extends figure~\ref{figIJ} to which it reduces upon taking a slice of fixed $\Omega$, except that we have reduced the plot data by not presenting the entire stellar family to avoid the additional fold visible in figure~\ref{figIJ} that is irrelevant for the issue of the ridge and just makes its rendering more difficult.

The advantage of a ridge in such a three-dimensional plot is that, in the presence of noise in the data, it will be easier to 
identify such an extended structure than a simple kink in the mass-radius diagram. On the negative side, much more data 
will be necessary, including the measurement of angular momenta which is not obvious from the radio pulses. 

\begin{figure*}[h!]
\includegraphics[width=0.7\columnwidth]{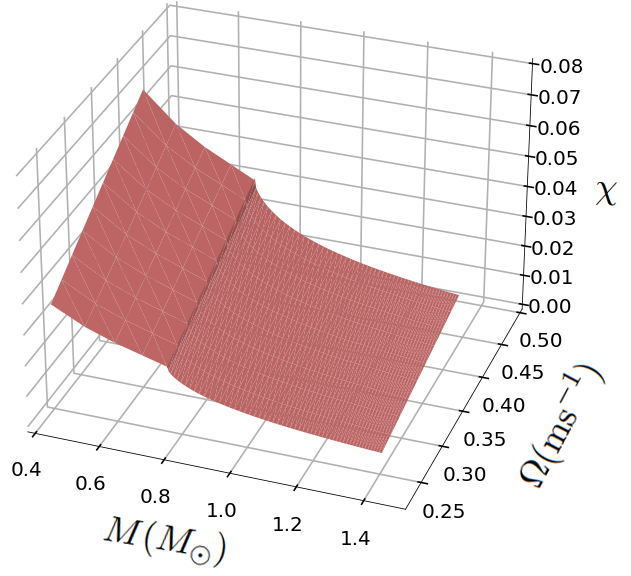}
\includegraphics[width=0.7\columnwidth]{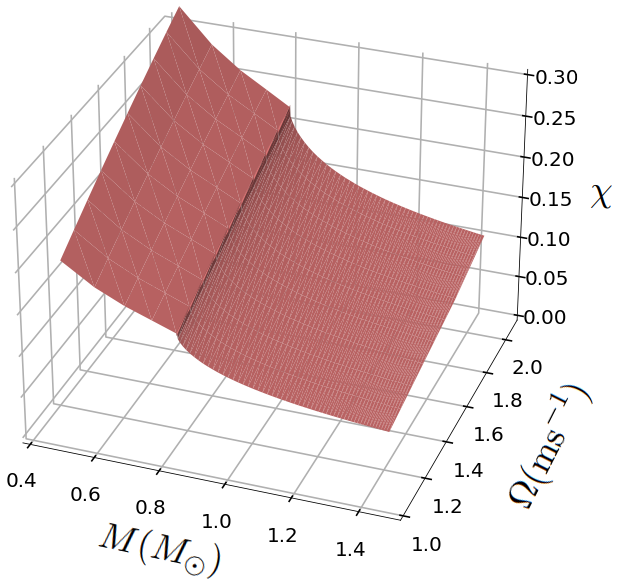}
\caption{\label{Ridgeinchi} Same as figure~\ref{RidgeinJ} but for the
adimensional angular momentum $\chi$ accessible for example through gravitational radiation in binary mergers.}
\end{figure*}

This brings us to figure~\ref{Ridgeinchi}, that transforms the variable $J$ to the adimensional angular momentum $\chi$ discussed around  Eq.~(\ref{chi_eff}).

\begin{figure*}[h!]
\includegraphics[width=0.7\columnwidth]{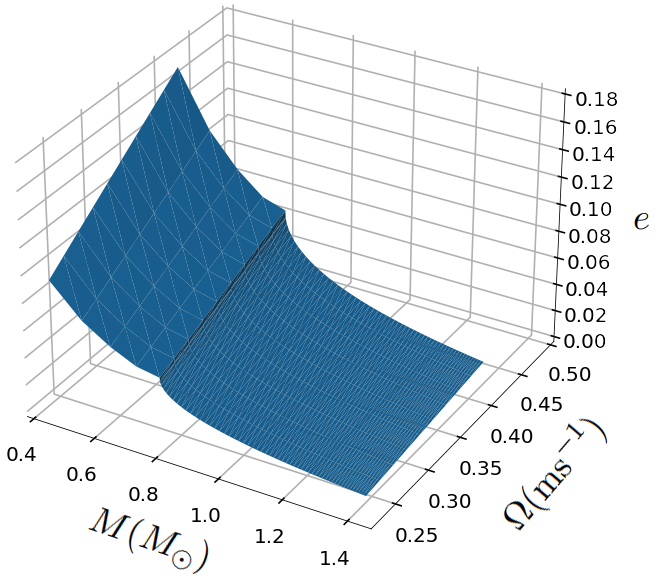}
\includegraphics[width=0.7\columnwidth]{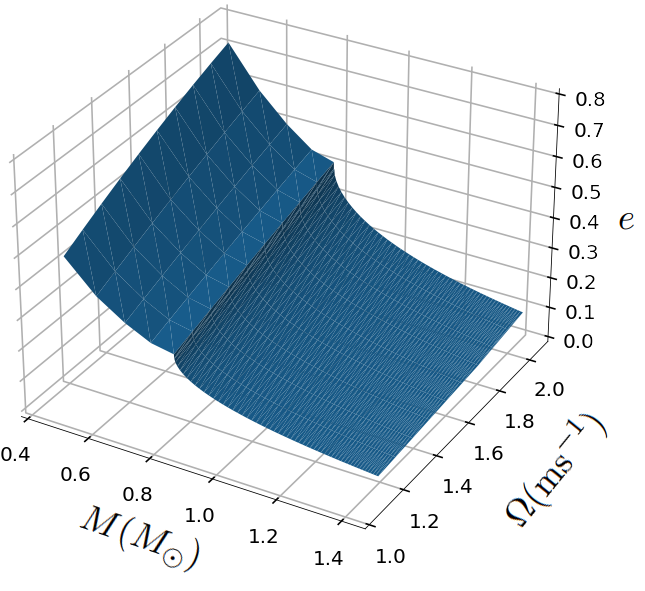}
\caption{\label{RidgeinEllipticity} The deformation of a rotating neutron star,
characterized by an ellipticity, displays a clear ridge when a star family is plot
against mass and angular velocity, if the EoS features a first order phase transition.}
\end{figure*}

Constraints from gravitational wave observatories can be brought to bear on $\chi$ and 
therefore the graph can one day be populated with observational data.  The same ridges
are visible, due to the artificially fed phase transition.

Finally, in figure~\ref{RidgeinEllipticity} we present a similar ridge caused by such a first order phase transition
in a plot of the ellipticity against the mass and the angular velocity.

This can be useful for X-ray observatories that are extracting radii and may eventually differentiate between 
polar and equatorial radii.

\clearpage
\section{Light rings or photon spheres around rotating neutron stars} \label{sec:lightring}

A longstanding prediction of General Relativity is that a circular $\dot{r}=0$
photon orbit at $r=3M$ should be present in the Schwarzschild geometry just outside
the Schwarzschild radius at $r=2M$. This feature has been claimed to be experimentally verified~\cite{Broderick:2022tfu}
for the black hole $M87^*$ by the Event Horizon Telescope's collaboration,
though the claim is still under discussion~\cite{Lockhart:2022rui}.

\begin{figure}
\includegraphics[width=0.9\columnwidth]{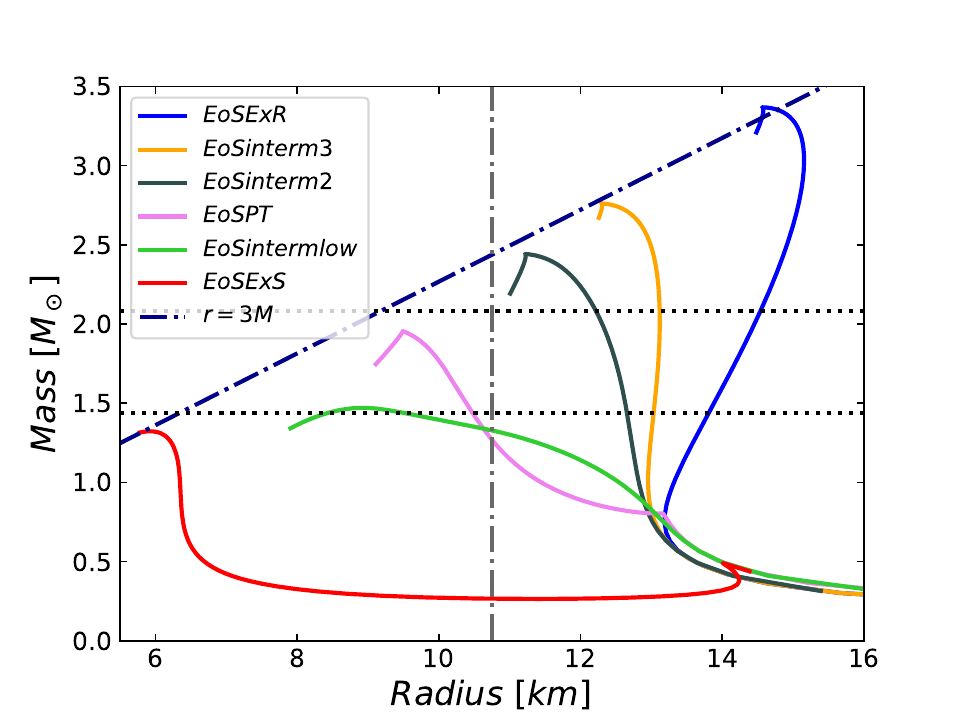}
\caption{\label{fig:RingofStatic}
The photon ring of the Schwarzschild geometry at $R=3M$ (corresponding to 6.2 km for a canonical 1.4 $M_\odot$ neutron star and 
to 10.1 km for an extreme 2.3 $M_\odot$ one) is not outside the star with typical equations of state, only the softmost and stiffmost ones 
of the nEoS band yield compact enough objects to feature such ring. But these EoS are disfavored within GR anyway and if used with 
modified gravity, the position of the ring should be recalculated anyway. 
See figure~\ref{fig_6EoS} in appendix~\ref{app:eos} for the actual equations of state used.
}
\end{figure}

The next to most compact objects known in relativistic astrophysics are neutron stars. 
In our appreciation, they are probably not compact enough to be able to support the photon ring
of the Schwarzschild geometry (see figure~\ref{fig:RingofStatic}). 
Only extreme equations of state within the nEoS band reach the line $M=R/3$ in the mass-radius diagram (remember the additional
1.47 kilometer per solar mass conversion factor). 
Since those EoS yielding the extremes of the band are probably excluded by astrophysical observables within
General Relativity, the fact that those at the ends do reach the line at $r=3M$ should rather be
reevaluated within modified gravity theories (see {\it e.g.} \cite{Astashenok:2020qds} for  rotating neutron stars in $f(R)$ theories and~\cite{daSilva:2022ctb} for $f(R,T)$).
This is confirmed by cursory examination of the mass-radius diagram in~\cite{Musolino:2023edi}. The upper end of the 
diagram with currently allowed equations of state (including astrophysical constraints) has a radius that is 
too large by about a kilometer, or 8\%, to be able to support a light ring.

Indeed, even without specific microscopic information about the equation of state, it has been known that in the so--called
Maximum Compact Configurations~\cite{Lattimer:2012nd}, that the radius had a lower bound $R>2.824\ M_\odot$
quite close to the photon ring position. This configuration is characterized by an EoS which is ultrasoft ($c_s^2=0$) at low
density and maximally stiff  ($c_s^2=1$) at high density. Such EoS are basically the opposite of what current hadron physics suggests, 
with an initially stiff regime driven by nucleon-nucleon repulsion as described by chiral forces, and an asymptotically softer
EoS at high densities due to QCD being conformal. Thus, it is not surprising that hadron--physics constraints push the EoS 
beyond the 3$M_\odot$ mark and the existence of a light ring for static stars does not seem viable.

\begin{figure}
\includegraphics[width=0.8\columnwidth]{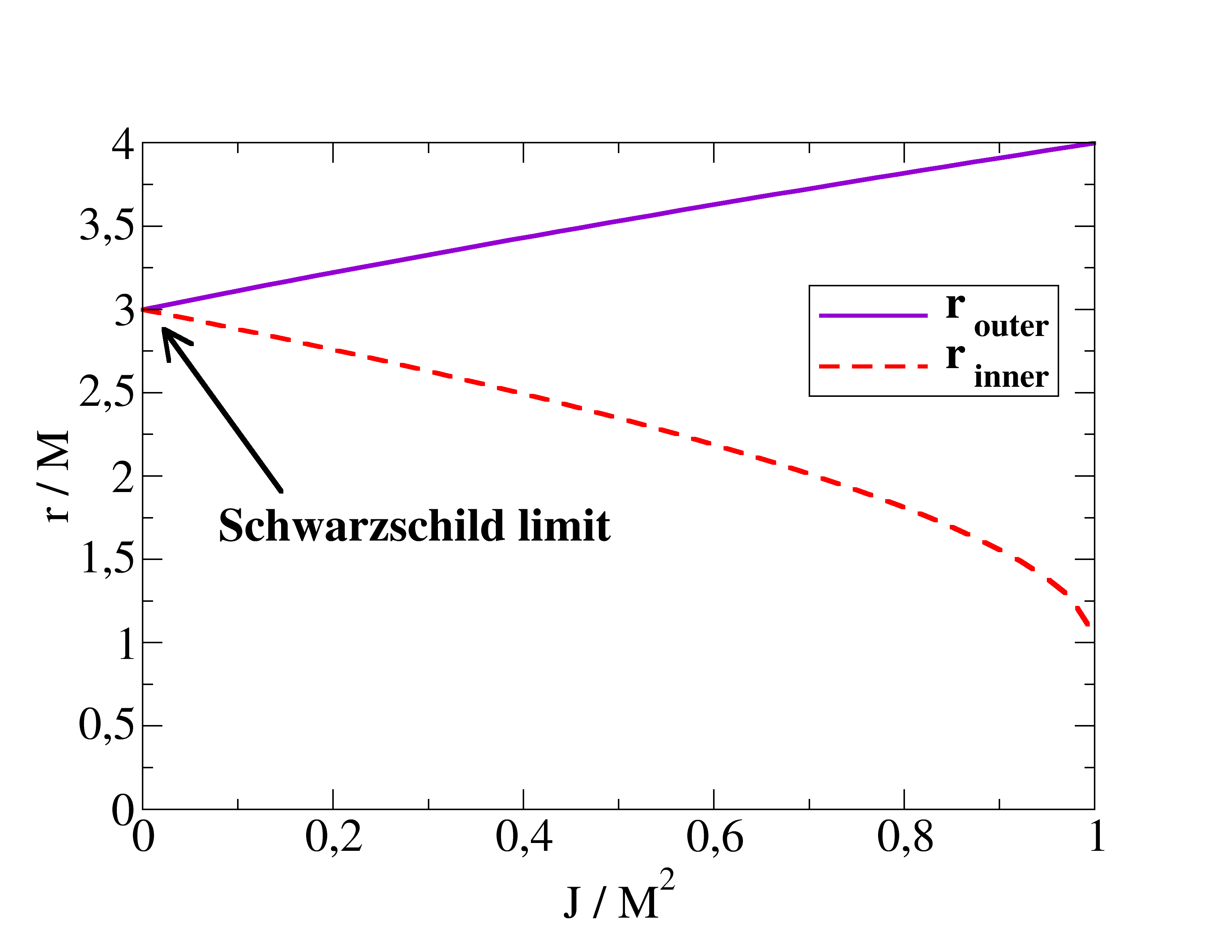}
\caption{\label{fig:RingofJ}
 Radial distance of the $\dot{r}=0$ photon orbits as function of the angular momentum (both normalized by the adequate 
power of the star mass). At $J=0$ both merge into the Schwarzschild photon ring. The outer photon sphere, due to the star's rotation, 
is further from the star even in proportion to the mass as the angular momentum increases. 
}
\end{figure}

For rotating matter configurations, two new phenomena are relevant~\cite{2003GReGr..35.1909T}. 
The first is that the light ring splits into two (in analogy with the two horizons of the Kerr metric), with radial coordinates
\begin{equation} \label{photonsphere}
r_\pm = 2 M \left(1+\cos\left( \frac{2}{3} {\rm arccos}\left(\pm \frac{|J|}{M^2}\right)
\right)\right).
\end{equation}
The inner radius is out of the question, but the outer radius increases with the angular momentum as shown in figure~\ref{fig:RingofJ}.
This may put it outside the star for EoS that are stiff but still allowed by other constraints, even in General Relativity. This is shown in figure~\ref{fig:RingofJ}
The second phenomenon is that the photon ring actually becomes a photon sphere~\cite{2003GReGr..35.1909T} 
with complicated photon orbits $(\theta(t),\varphi(t))$ at fixed $r$.

\begin{figure}
\includegraphics[width=0.8\columnwidth]{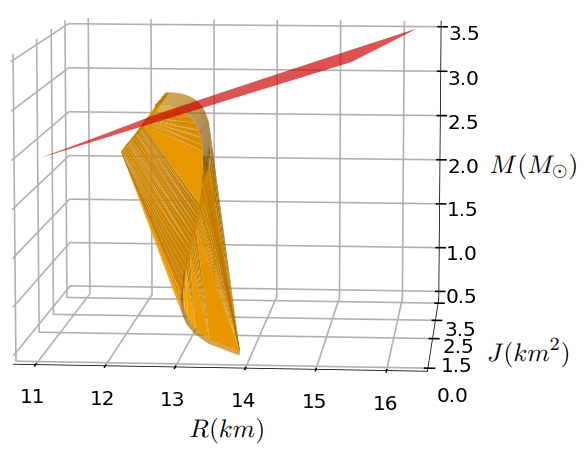}
\includegraphics[width=0.8\columnwidth]{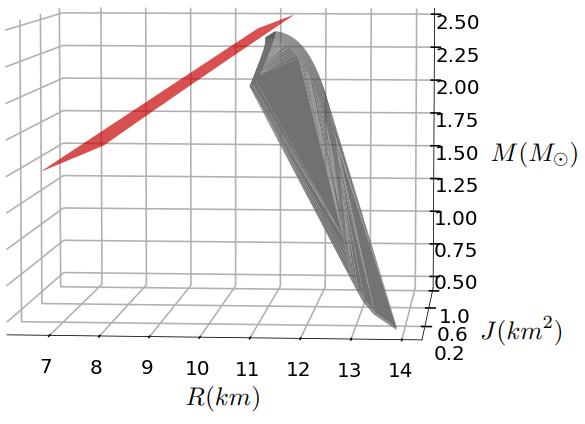}
\caption{\label{fig:RingofJ2} 
Rotating stars in the presence of an even moderately stiff equation of state may feature a photon sphere,
as manifested by the intersection of two surfaces, the first being the condition of Eq.~(\ref{photonsphere}) for $r_+$
and the second the $M(R,J)$ diagram that extends the traditional $M(R)$ diagram to rotating stars.
The top plot corresponds to the stiffer EoS Interm3 from figure~\ref{photonsphere} whereas the bottom plot
to the softer Interm2 (with a first order phase transition that turns the mass-radius diagram in figure~\ref{photonsphere}).
The top one would entail a photon sphere for the heaviest neutron stars, the bottom plot the absence thereof for all stars. 
The existence of such photon sphere/ring is seen to be a possible discriminant for phase transitions in neutron star matter.
}
\end{figure}

It is apparent from figure~\ref{fig:RingofJ2} that rotating stars can definitely feature photon spheres with reasonable 
still allowed EoS. But since the margin is not so large, we envision that, should such photon spheres/rings ever be found
in a very massive neutron star, they would put into question any first order phase transition with reasonably large latent heat,
as the star of characteristic typical radius would not reach the necessary mass.

\section{Outlook}\label{sec:outlook}

While true that detailed investigations of rotating stars in the presence of an exotic phase of neutron matter~\cite{Rather:2021yxo,Bandyopadhyay:2017dvi}
have been reported, our focus has been, rather than in attempting to provide a realistic description of the overall 
set of observables, to provide credible target phenomena for the most interesting searches. These aim, in our opinion, to find the first order phase transition with a finite latent
heat~\cite{Lope-Oter:2021mjp} that forces a non smooth behaviour in the various diagrams that relate observable to observable; 
this is in analogy to the kinks 
long predicted in the mass-radius or mass-moment of inertia diagram, and we think can be of future use.

The kink in the $I(M)$ diagram requires measuring, for the same star, the angular momentum $J$ and 
 frequency $\Omega$ to obtain their ratio, compounding the uncertainties. This difficults the identification of the feature
in the diagram until exquisite data is at hand. Unfolding the two variables in a three-dimensional diagram makes the uncertainties
stretch in different directions, forming an ellipse in the XY base plane of the diagram; allows to have an entire line of such
kinks to be identified; and offers the possibility of having lines where the kink is more pronounced (and easier to identify)
rather than having only an average of the best and worst cases in a two-dimensional diagram. On the down side, more data will be
necessary to populate the diagram to some given density of points.

Although we have employed the slow-rotation expansion in powers of $\Omega R/c$, it appears that 
our results are relevant for all known pulsars, except perhaps PSR J1748-2446ad that, at 716 Hz spinning frequency~\cite{Hessels:2006ze},
may have a surface rotating at 15-20\% of  light speed and corrections may therefore be large. 
Further insight in phase transitions can be gained
from such rapidly rotating objects~\cite{Harko:2004zz}, whether isolated such as this one
or more likely in the dynamic setting of a binary collision, 
and that requires further theory (and numerical simulation).

Another issue that we have not confronted is that of systematic statistical analysis to look for phase transitions~\cite{Mroczek:2023eff,Musolino:2023edi} or systematically constrain the overall shape of the EoS~\cite{Raaijmakers:2021uju,Volkel:2022utc}, as we do use the nEoS hadron-theory band but have not attempted to thoroughly sample it. This is off the point given that the sampling will only move the position of the eventual nonanalyticity in the various diagrams, but not erase it if a first order phase transition is present in neutron matter. There is no difficulty of principle in attempting a realistic mapping of the possible position of such singularity along the lines
of~\cite{Providencia:2023rxc}
and interesting future work could proceed along those lines. 

Finally, we have ventured to present, in exploratory form, a counterobservable whose finding would suggest the opposite,
the absence of a phase transition: this is the photon sphere around a rotating star. Because of their faintness,
these will be hard to search for, but methods will hopefully be devised in this very active field.

\acknowledgments
We thank Micaela Oertel for providing useful comments to the first preprint.
Work partially supported by the EU under grant 824093 (STRONG2020); 
spanish MICINN under PID2019-108655GB-I00, PID2019-106080GB-C21; 
Univ. Complutense de Madrid under research group 910309 and the IPARCOS institute. 

This preprint is issued with numbers \\ IPARCOS-UCM-23-049 and ET-0262A-23



\appendix  
\section{Brief overview of the Hartle-Thorne rotating star structure equations}
\label{app:rotation}

Because the metric in Eq.~(\ref{metrica2}) is invariant under simultaneous time and azimuth flip,
$t\longrightarrow -t$,  $\phi\longrightarrow -\phi$, only even powers of $\Omega$ appear upon expanding
 $H,Q$ and $K$, while $L$ will be odd. Thus, to first order, $L$ will be a function that can appropriately be named $\omega$ as it is linear in $\Omega$, $ L(r,\theta)=\omega(r,\theta)+O(\Omega^3)$.

The constitutive stellar equations are written in terms of $\varpi\equiv\Omega-L$, the difference between the reference 
``geometric'' angular velocity $\Omega$ as seen from infinity and $L(r,\theta)\sim\omega$, that of an observer at a local inertial reference frame that is being dragged. 

The four-velocity of the rotating fluid, necessary to construct the ideal energy-stress tensor $T^{\mu\nu}=(\rho+P)u^\mu u^\nu + Pg^{\mu\nu}$,
has both temporal and azimuthal components
\begin{eqnarray}\label{umu}
    u^{\mu}&=&(u^t,0,0,\Omega u^{t}),\\
    u^t&=&(-g_{tt}-2\Omega g_{t\phi} -\Omega^2g_{\phi\phi})^{-1/2}
\end{eqnarray}
that can be computed from the interval tensor $g$, guaranteeing $u^2=-1$.

\begin{figure}[h!]
\centering
\begin{tikzpicture}
        \draw circle(1.8);
        \draw [dashed] circle(2.2);
        \draw (0,0) -- (0.9,0) node[above] {$R$} -- (1.8,0);
        \draw[very thick, <-] (1.8,0) -- (2.2,0);
        \draw[very thick, <-] (2.2,0) -- (1.8,0);
        \draw (2.2,0) node[right] {$\xi_0$};       
    \end{tikzpicture}\\ \vspace{0.3cm}
    \begin{tikzpicture}
        \draw circle(1.8);
        \draw [dashed] (0,0) ellipse (2.5 and 1.4);
        \draw (0,0) -- (0.9,0) node[above] {$R$} -- (1.8,0);
        \draw[very thick, <-] (1.8,0) -- (2.5,0);
        \draw[very thick, <-] (2.5,0) -- (1.8,0);
        \draw (2.4,0) node[right] {$-\xi_2/2$};    
    \end{tikzpicture}
    \caption{Spherical  $l=0$ and quadrupolar $l=2$ deformations, the 
    second one yielding the ellipticity.}
    \label{deformations}
\end{figure}
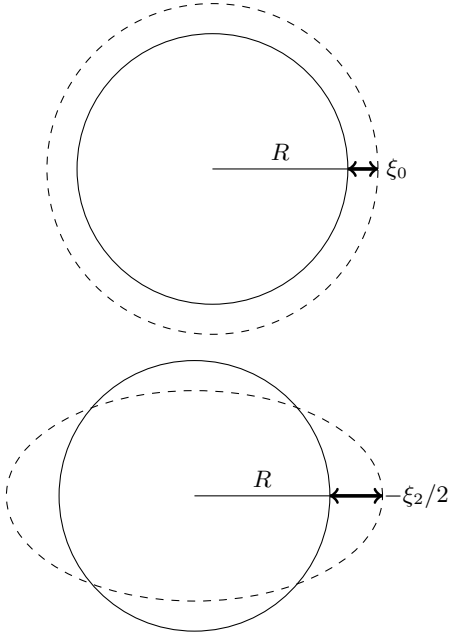

Then the angular velocity relative to the surface, $\varpi$, is calculated from Einstein's time-azimuthal equation
\begin{eqnarray}\label{Rphit}
    R_{\phi}^{\;\;t}=8\pi T_{\phi}^{\;\;t}. \\
\label{eq.varpi}
  \frac{\partial \!\left(\!r^4j\varpi_{,r}\!\right)}{r^4\partial r}\!+\!\frac{4\varpi}{r}j_{,r}+\frac{e^{(\lambda-\nu)/2}}{r^2}\frac{1}{\sin^3\!\theta}\frac{\partial}{\partial\theta}\!\left(\!\sin^3\!\theta \varpi_{,\theta}\!\right)\!=0,
\end{eqnarray}
where an auxiliary adimensional quantity has been introduced,
\begin{equation}\label{j}
    j(r)=e^{-(\lambda+\nu)/2}=e^{-\nu/2}\sqrt{1-2m/r},
\end{equation}
that in the Schwarzschild static limit is trivial $j(r)=1$. The radial derivative of Eq.~(\ref{j}), upon substituting the static TOV system and employing again Eq.~ (\ref{j}), can be obtained from 
\begin{equation}\label{derj}
    -\frac{dj}{dr}=4\pi r(\rho+p)\frac{e^{-\nu/2}}{\sqrt{1-2m/r}}=4\pi r(\rho+p)\frac{e^{-\nu}}{j}.
\end{equation}
Equation~(\ref{eq.varpi}) is treated by separating variables and expanding the polar dependence in $\varpi(r,\theta)$ via vector spherical harmonics, which returns an expansion in terms of Legendre polynomials
\begin{equation}\label{des.varpi}
    \varpi(r,\theta)=\sum_{l=1}^{\infty}\varpi_l(r)\left(-\frac{1}{\sin{\theta}}\frac{dP_l}{d\theta}\right)\ .
\end{equation}
That would yield, in general, the differential rate of rotation of rings along latitude parallels, with each $l$ component satisfying
\begin{equation}\label{eq.varpi2}
    \frac{1}{r^4}\frac{d}{d r}\left(r^4j\frac{d\varpi_l}{dr}\right)+\left[\frac{4}{r}\frac{dj}{dr}-e^{(\lambda-\nu)/2}\frac{l(l+1)-2}{r^2}\right]\varpi_l=0\ .
\end{equation}
Avoiding singularities at $r\to \infty$ where planar geometry must be recovered, and at $r\to 0$
greatly constrains their possible values, with all terms vanishing save $l=1$ that yields the assymptotic value of the angular velocity
 $\varpi_1(r)\longrightarrow b_2=\Omega$;
 $\varpi=\varpi_1$ becomes a purely radial function so that the subindex can be dropped, and the rotation of the star does not depend on the latitude. The equation for $dj/dr$ is then simpler,
\begin{equation}\label{eq.varpi3}
    \frac{d}{dr}\left(r^4j(r)\frac{d\varpi}{dr}\right)+4r^3\frac{dj}{dr}\varpi=0\ .
\end{equation}
If the zeroth order of the interval in  Eq.~(\ref{metrica2}) is extracted, taking into account that such $\Omega^0$ order must reproduce the Schwarzschild metric of Eq.~(\ref{metrica}), we  can rewrite it in terms of functions $h(r,\theta)$, $m(r,\theta)$ and $k(r,\theta)$ that are second order in $\Omega$
\begin{eqnarray}\label{metrica3} \nonumber 
    ds^2 &=& -e^{\nu}(1+2h)dt^2 +  e^{\lambda}\left[1+\frac{2m}{r-2M}\right]dr^2
     \\ &+& r^2(1+2k)(d\theta^2+\sin^2{\theta}(d\phi-\omega dt)^2)\ .
\end{eqnarray}
These functions have an expansion in spherical harmonics from which we keep also the second order
\begin{eqnarray}\label{h}
    h(r,\theta)=h_0(r)+h_2(r)P_2(\theta)+...
\\ \label{m}
    m(r,\theta)=m_0(r)+m_2(r)P_2(\theta)+...
\\  \label{k}
    k(r,\theta)=k_0(r)+k_2(r)P_2(\theta)+...
\end{eqnarray}
By choosing the $r$ variable appropriately, $k_0(r)=0$ is conventionally fixed, and $k_2$ is traded for the variable
 $v_2=h_2+k_2$. 
Solving Einstein's equations for the $l=0$ and $l=2$ perturbations~\cite{Hartle:1968si,Posada:2023bnm} (that are not coupled and can be separately treated) yields the spherical and quadrupolar star deformations shown in figure~\ref{deformations}.

The isobaric $P=0$ surface delimiting the star
\begin{equation}\label{rp}
    r_p(\theta)=R+\xi_0(R)+\xi_2(R)P_2(\cos{\theta})
\end{equation}
is parametrized by two auxiliary variables $p_0^*$ and $p_2^*$ that absorb some additional functions,
\begin{eqnarray}\label{xi0}
    \xi_0(r)=-p_0^*(\rho+p)\left(\frac{dp}{dr}\right)^{-1},
\\\label{eq.xi2}
    \xi_2(r)=-p_2^*(\rho+p)\left(\frac{dp}{dr}\right)^{-1}.
\end{eqnarray}

\subsection{Set of equations for $l=0$ spherical deformation}
The resulting equations for the constitutive $l=0$ functions are then as follows:
\begin{equation}\label{eq.l01}
    \frac{d}{dr}\left(r^4j(r)\frac{d\varpi}{dr}\right)+4r^3\frac{dj}{dr}\varpi=0,
\end{equation}
\begin{equation}\label{eq.l02}
    \frac{dm_0}{dr}=4\pi r^2(\rho+p)\left(\frac{d\rho}{dp}\right)p_0^*+\frac{1}{12}r^4j^2\left(\frac{d\varpi}{dr}\right)^2-\frac{1}{3}r^3\varpi^2\frac{dj^2}{dr},
\end{equation}
\begin{eqnarray}\label{eql03}
    \frac{dp_0^*}{dr} &=& -\frac{1+8\pi pr^2}{(r-2m)^2}m_0 - \frac{4\pi(\rho+p)r^2}{r-2m}p_0^* \nonumber\\ &+&\frac{1}{12}\frac{r^4j^2}{r-2m}\left(\frac{d\varpi}{dr}\right)^2+\frac{1}{3}\frac{d}{dr}\left(\frac{r^3j^2\varpi^2}{r-2m}\right). \nonumber \\
\end{eqnarray}
They form an ordinary  Cauchy system linear on the highest derivatives and are integrated from the stellar center ($r=0$) outwards with initial conditions  $m_0(0)=0$, $p_0^*(0)=0$, $\varpi(0)=\varpi_c$ and $\displaystyle\left.\frac{d\varpi}{dr}\right|_{r=0}=0$. 
The pressure profile is then technically a perturbation of that for a static star of equal central pressure. We conveniently choose
 $\varpi_c=1$, which then determines the outcome for the angular velocity at the star's surface, $\Omega$. Since this is the parameter that
 tags the solutions and is observable from infinity, we would like to be able to choose it in the calculation; this is possible thanks to the universal relation~\cite{Hartle:1968si,Glendenning}
\begin{equation}\label{RelacionOmega}
    \frac{\varpi(r)}{\Omega}=\frac{\varpi'(r)}{\Omega'}.
\end{equation}
That follows from the linearity of Eq.~(\ref{eq.l01}) for $\varpi(r)$ and is fed its value at the star's surface as shown shortly.
To numerically integrate Eq.~(\ref{j}) and (\ref{derj}) and Eq.~(\ref{eq.l01}) we double this to a first order system and then employ the fourth order Runge-Kutta method. The actual system being put on the computer is thus

\begin{eqnarray}\label{eq.alpha}
    \frac{d\varpi}{dr}=\bar\alpha,
\\ \label{eq.deralpha}
    -\frac{d\bar\alpha}{dr}=\frac{1}{j}\frac{dj}{dr}\left[\bar\alpha+\frac{4}{r}\varpi\right]+\frac{4}{r}\bar\alpha,
\\ \label{eq.m0}
    \frac{dm_0}{dr}=4\pi r^2(\rho+p)\left(\frac{d\rho}{dp}\right)p_0^* \nonumber \\ +\frac{1}{12}r^4j^2\left(\frac{d\varpi}{dr}\right)^2-\frac{2}{3}r^3\varpi^2j\frac{dj}{dr},
\\ \label{eq.p0}
    \frac{dp_0^*}{dr}=-\frac{1+8\pi pr^2}{(r-2m)^2}m_0-\frac{4\pi(\rho+p)r^2}{r-2m}p_0^* 
    \nonumber \\  +\frac{1}{12}\frac{r^4j^2}{r-2m}\left(\frac{d\varpi}{dr}\right)^2+\frac{1}{3}K(r),
\\ \label{eq.K}
    K(r)=\frac{r^3j^2\varpi^2}{(r-2m)^2} \nonumber \times \\ \left[(r-2m)\left(\frac{3}{r}+\frac{2}{j}\frac{dj}{dr}+\frac{2}{\varpi}\frac{d\varpi}{dr}\right)+2\frac{dm}{dr}-1\right].
\end{eqnarray}
Integration of Eq.~(\ref{eq.m0}) outside the star allows to match at infinity and define the mass increase due to the rotation, $\delta m$, 
\begin{equation}\label{deltaM}
    \delta m=m_0(R)+\frac{J^2}{R^3},
\end{equation}
in terms of the total angular momentum $J$.

\begin{figure}[h]
\centering
\includegraphics[width=\linewidth]{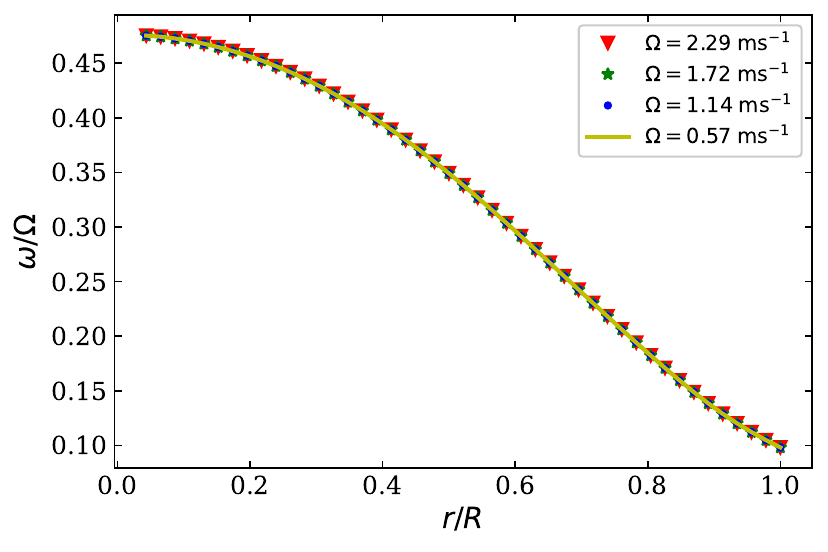}
\includegraphics[width=\linewidth]{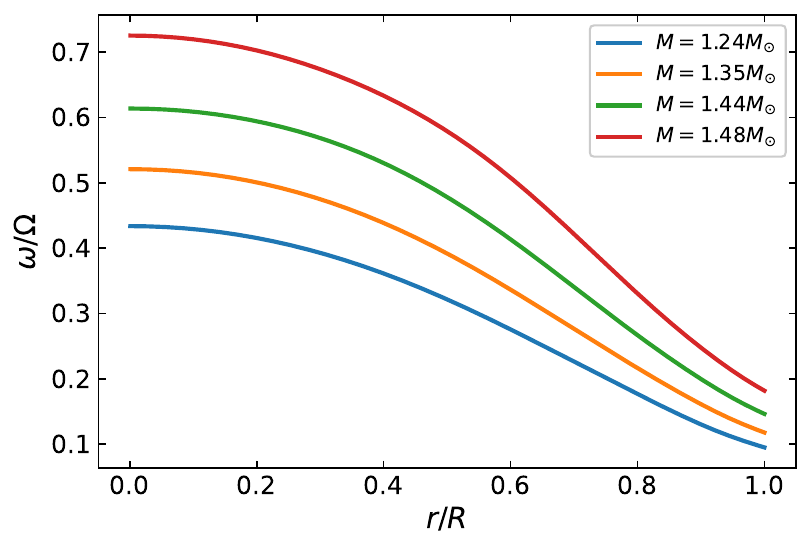}
\caption{
Angular velocity profile in a stellar interior. Top: universal profile for a fixed star of mass  $M=1.3\;M_{\odot}$ for different reference angular velocities $\Omega$.
Bottom: rescaled profiles for different stars distinguished by their total mass $M(R)$. 
}
\label{w_profile}
\end{figure}

In figure~(\ref{w_profile}) we show the resulting angular velocity profile inside the star
(not a direct observable, but a needed auxiliary quantity). The equation of state employed is the intermediate one 
from figure~\ref{eosfig} (blue line). 
The top plot demonstrates the validity of the rescaling of Eq.~(\ref{RelacionOmega}) that can also be taken as a check of the computer code:
the profiles corresponding to different tagging values $\Omega$ at the surface perfectly fall on each other after dividing $\omega(r)$ by $\Omega$.
The bottom plot then shows $\omega(r)/\Omega$ for several different stars near the canonical one for the resulting family.
Our numerical results seem to be in agreement with those from the Barcelona group~\cite{Campoy}.

Both plots clearly show that the local drag angular velocity (remember that this is the one that controls the local inertial forces)
$\omega$, monotonically decreases outwards from the stellar center in a nontrivial manner: rotation is not rigid, yet the fluid is ideal, there are no viscous forces nor dissipation among different layers of the star, the effect being purely relativistic.
It is also clear that $\omega< \Omega$ for the entirety of the star's interior.

\subsection{Set of equations for the $l=2$ quadrupolar deformation}
The equations for the constituting  $l=2$ functions are then as follows:
\begin{eqnarray}\label{eq.h2}
    \frac{dh_2}{dr}=\!\left\{\!\frac{r}{r-2m}\!\left(\!\frac{d\nu}{dr}\!\right)^{-1}\!\!\left[\!8\pi(\rho\!+\!p)-\frac{4m}{r^3}\!\right]
    -\frac{d\nu}{dr} \right\}h_2 \nonumber \\ -\frac{4v_2}{r(r-2m)}\left(\frac{d\nu}{dr}\right)^{-1} \nonumber \\
   + \frac{1}{6}\left[\frac{r}{2}\left(\frac{d\nu}{dr}\right)-\frac{1}{r-2m}\left(\frac{d\nu}{dr}\right)^{-1}\right]r^3j^2\left(\frac{d\varpi}{dr}\right)^2 \nonumber \\ -\frac{1}{3}\left[\frac{r}{2}\left(\frac{d\nu}{dr}\right)+\frac{1}{r-2m}\left(\frac{d\nu}{dr}\right)^{-1}\right]r^2\frac{dj^2}{dr}\varpi^2, \\
\label{eq.v2}
    \frac{dv_2}{dr}=-\left(\frac{d\nu}{dr}\right)h_2\nonumber \\  +\left(\frac{1}{r}+\frac{1}{2}\frac{d\nu}{dr}\right)\left[\frac{1}{6}r^4j^2\left(\frac{d\varpi}{dr}\right)^2-\frac{1}{3}r^3\varpi^2\frac{dj^2}{dr}\right]\ .
\end{eqnarray}
This is again a Cauchy system with initial conditions $h_2=v_2=0$ at $r=0$ and $r\longrightarrow\infty$ and that takes as input the solution to the static TOV system. 
Once integrated, the quadrupolar perturbations to the mass distribution $m_2$ and to the stellar surface $p_2^*$ follow from
\begin{eqnarray}\label{eq.m2}
    m_2(r)&=&(r\!-\!2m)\left[\frac{1}{6}r^4j^2\left(\frac{d\varpi}{dr}\right)^2 -h_2-\frac{2}{3}r^3j\frac{dj}{dr}\varpi^2  \right]\nonumber \\ 
\\ \label{eq.p2}
    p_2^*(r)&=&-h_2-\frac{1}{3}r^2e^{-\nu}\varpi^2\ .
\end{eqnarray}

\section{Newtonian limit}
As a check,  the nonrelativistic limit  $p\ll\rho$ and $2m/r\ll 1$ in equations~(\ref{TOV1}) and~(\ref{TOV2})
recovers the classical equations of nonrelativistic hydrostatic  equilibrium
\begin{eqnarray}\label{classicTOV1}
    \frac{dM(r)}{dr}=4\pi r^2\rho(r),
\\ \label{classicTOV2}
    -\frac{dp(r)}{dr}=\frac{\rho(r)M(r)}{r^2}\ .
\end{eqnarray}
The gravity source is of course the star's density alone $\rho(r)$, giving Poisson's equation
\begin{equation}\label{poisson}
    \Delta\Phi(r)=4\pi\rho(r)\ .
\end{equation}

Proceeding to a stationary rotating star, the corrections~\cite{Hartle:1967he,Boshkayev} to the static one, within Newtonian mechanics,
are given by

\begin{eqnarray}
    \frac{dM^{[2]}(r)}{dr}&=&4\pi r^2\frac{d\rho}{dp}\rho p^*_0 \\
    -\frac{dp^*_0(r)}{dr}+\frac{2}{3}\Omega^2r&=&\frac{M^{[2]}(r)}{r^2}
\\
    \delta M&=&M^{[2]}(R)\ .
\end{eqnarray}

The stellar surface is then obtainable from
\begin{equation}\label{def_newt}
    r=R+\xi_0(R)+\xi_2(R)P_2(\theta),
\end{equation}
where
\begin{eqnarray}\label{xi_0_new}
    \xi_0(R)=\frac{R^2}{M}p_0^*(R)
\\\label{xi_2_new}
    \xi_2(R)=-\frac{R^2}{M}\left(\frac{1}{3}\Omega^2R^2+\Phi_2^{[2]}(R)\right)\ .
\end{eqnarray}
The term $\Phi_2^{[2]}(R)$ appearing in Eq.~(\ref{xi_2_new}) corresponds to the second term in the  $\Omega$ expansion of $\Phi(r,\theta)$
\begin{equation}
    \Phi(r,\theta)=\Phi^{[0]}(r)+\Phi^{[2]}(r,\theta)+O(\Omega^4)\ ,
\end{equation}
with
\begin{equation}
    \Phi^{[2]}(r,\theta)=\sum_{l=0}^{\infty}\Phi^{[2]}_l(r)P_l(\cos{\theta})\ .
\end{equation}
Finally, the ellipticity can be computed by means of Eq.~(\ref{e}), where $r_{\rm{Polar}}$ and $r_{\rm{Equatorial}}$ are the same as in Eq.~ (\ref{rpolar}) and~(\ref{req}).

The angular momentum computed for a rotating distribution with rigid angular velocity $\Omega$ is easily computed to be, 
in Newtonian mechanics,
\begin{equation}\label{Jnew}
    J_{Newton}=\frac{8\pi}{3}\Omega_c\int_0^R dr r^4\rho\ .
\end{equation}

\begin{figure}
\includegraphics[width=0.8\columnwidth]{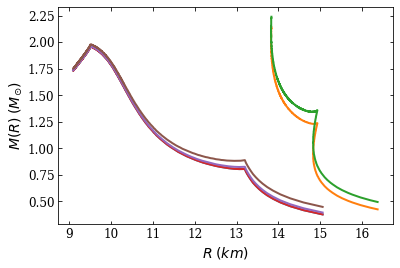}
\caption{\label{MRNewton} Mass-radius diagram in the Hartle-Thorne relativistic theory (curves to the left, yielding more compact stars)
and in Newtonian mechanics (those to the right, with larger radii). The presence of kinks due to the underlying phase transition in the equation of state is independent of the theory of gravity.}
\end{figure}

A numerical evaluation of the Newtonian constitutive equations with the same equations of state and angular velocities as in General Relativity allows comparing the classical nonrelativistic solutions with  the relativistic Hartle--Thorne ones. 
Doing so for the orange intermediate nEoS in figure~\ref{eosfig} yields the mass-radius diagram shown in figure~\ref{MRNewton}. 
The relativistic and nonrelativistic curves converge for sparse stars at large radius and small mass on the bottom right corner of the plot. 

For more compact stars, the Newtonian ones are larger (gravity is less intense, there not being a Schwarzschild radius,
and pressure does not gravitate) but the shape is similar, and to the point of this work,
the discontinuity in the derivative  is observed in both the Newtonian and the relativistic star families.
 This is of course in line with our expectation that this feature is independent of the theory of gravity and can be used,
 if we were lucky enough that it would be present in the data, to start disentangling effects due to the equation of state from effects due to modifications of gravity, that often lead to degeneracy.

\clearpage

\section{Construction of the nEoS band of Equations of state and of the illustrative examples here employed} \label{app:eos}
In this appendix we review the construction of an EoS band based on first principles, specifically relying on the foundation of QCD, 
its low--density effective field theory, and basic theoretical principles. The philosophy of the work can be found in the original
reference~\cite{Oter:2019rqp}, whose sets can be freely downloaded from~{\tt http://teorica.fis.ucm.es/nEoS} and further
refinements have been presented in~\cite{Lope-Oter:2021vxl,Lope-Oter:2021mjp,Lope-Oter:2023urz}.

One usually speaks of  ``stiff'' or ``soft'' Equations of State. Generally speaking, a stiff EoS has large slope in the pressure-energy density diagram $P(\epsilon)$ ($\epsilon=\rho c^2$ in this article), even saturating 
causality near $c_s^2\lesssim 1$,  and the radius of the neutron star increases (or remain roughly constant in certain regions of the mass-radius diagram) as the mass increases. In other words, the pressure inside the compact object is sufficiently high to counteract the gravitational attraction. On the contrary, a soft EoS will have smaller slope $P'(\epsilon)$ which leads to a decrease in the star's radius as the mass increases and is more compressed. Phase transitions occur if the new, exotic phase relaxes the free energy, and therefore always entail some degree of softening, the more intense the larger the latent heat.

The nEoS band is organized in three density regimes, with the lowest and highest density controlled by Lagrangian--based nuclear and particle physics, and the intermediate one by basic theory assumptions only:
\begin{itemize}
    \item \textbf{Low ($\chi$EFT regime)}\\ 
At low densities $0.05 \leq n \leq 0.34$ fm$^{-3}$, we utilize the computed data from \cite{Drischler:2020hwi,Drischler:2020yad}. Since the $\chi$EFT EoS are provided in terms of $n$ (number density) and the binding energy per nucleon $E/A$, we extract the EoS (pressure $P$ as a function of the energy density $\varepsilon$) from their data using
\begin{equation}\label{eq:densityenergy}
\varepsilon= n\left(M_N+\frac{E}{A}\right)
\end{equation}
\begin{equation}\label{eq:firstlaw}
P= n^2\frac{d(E/A) }{d n},
\end{equation}  
where Eq.~(\ref{eq:densityenergy}) represents the expression for the energy density, including the rest mass of the particles, and Eq.~(\ref{eq:firstlaw}) is the first law of thermodynamics at $T=0$. 
Chiral Perturbation theory is not renormalizable, so the authors employ ultraviolet cutoffs set at $450$ and $500$ MeV in \cite{Drischler:2020hwi,Drischler:2020yad}, and we use the two calculations to obtain an uncertainty spread for this low-density region.
\item \textbf{Very high (pQCD regime)}\\
At asymptotically high--density, perturbative Quantum Chromodynamics is supposedly valid as $\alpha_s(\mu_B)$, the coupling constant 
at the large baryon chemical potential, should become a small parameter, and provide computable thermodynamic quantities with
small corrections due to possible gapped colour-superconducting phases.
For densities $n \geq 40 n_s$, we consider as a matching point with the pQCD region~\cite{Kurkela:2022elj} a specific baryon chemical potential value $\mu_B \approx 2.6$ GeV, within an uncertainty band between different values of the scale parameter\footnote{ The scale parameter $X\equiv 3\bar{\Lambda}/\mu_B$, where $\bar{\Lambda}$ is the renormalization scale.} $X$ defined by the nordic group~\cite{Gorda:2018gpy} ranging from $1$ to $4$. Similar to the $\chi$EFT region, we utilize the energy density, pressure, and number density values from~\cite{Gorda:2021znl} computed for these two values of $X$ to provide a reasonable systematic theoretical uncertainty band in this high-density regime.
  \item \textbf{Intermediate density with interpolated EoS} \\
In the intermediate region between the chiral domain and the pQCD regime, we close the maximum allowable region (Rhoades-Ruffini rhomboid) based on the conditions of causality ($c^2_s\leq 1$) and monotonic behavior ($c^2_s\geq 0$). Within this region, we construct a grid of candidate points ($\varepsilon, P$) through which potential EoS candidates may pass, following the interpolation procedure developed in \cite{Oter:2019rqp}. The EoS can be generated in two different ways: randomly \cite{Oter:2019rqp} for statistical sampling
or by controlling the slope $c_s^2$. The latter approach involves prolonguing the same growth rate of the slope from the chiral region into the intermediate interpolation zone, as far as the grid permits~\cite{Evaphd}, which is a common--sense guess with no better claim than 
any other EoS within the band (except that the so constructed EoS would additionally satisfy astrophysical mass constraints, whereas the band as a whole is astrophysics-agnostic to allow for exploring alternative theories of gravity). To accomplish this, we construct a 100-point EoS using a 1000 $\times$ 1000-point grid to achieve acceptable control over the slope. The specific details of this procedure, employed in the example EoS in this investigation, have been discussed in \cite{Oter:2019rqp,Evaphd}.
\end{itemize}

 In this work, we have used the later approach. We construct equations of state (EoS) based on the candidate points of the grid, taking into account causality and monotonicity and thermodynamic consistency by utilizing the following discrete equations:
\begin{eqnarray}
n_i &=& \frac{\varepsilon_i}{M_N+(E/A)i}\label{eq:discrenumberdensity}\\
P_i &=& n^2_i \frac{(E/A){i+1}-(E/A){i-1}}{n{i+1}-n_{i-1}}
\label{eq:discrefirstlaw}\\
\mu_{Bi} &=& \frac{\varepsilon_i+ P_i}{n_i},
\label{eq:Euler}
\end{eqnarray}
Here, equations Eq.~(\ref{eq:discrenumberdensity}) and Eq.~(\ref{eq:discrefirstlaw}) correspond to equations Eq.~(\ref{eq:densityenergy}) and Eq.~(\ref{eq:firstlaw}), respectively, expressed for the discrete points derived from the grid. Equation Eq.~(\ref{eq:Euler}) represents the Euler equation, which ensures thermodynamic consistency and links the $(\epsilon,P)$ and $(n,\mu_B)$ diagrams so that the uncertainty band in the usual EoS plane becomes a tube in a multidimensional space. We impose all constraints from microscopic physics known to us, in particular  the limiting values of ($\varepsilon, P, \mu_B, n$) obtained from the pQCD regime which, given the initial stiffness due to chiral effective theory, suggest a long first-order phase transition to reach the matching baryon chemical potential $\mu_B=2.6$ GeV without having violated any of them: these constraints derived from thermodynamic consistency lead to a softening of the EoS for high-density.

 In addition to providing information on energy density $\varepsilon$ and pressure $P$, our approach also includes details such as baryon density number $n$, binding energy per nucleon $E/A$, sound speed $c_s^2$ (representing the slope), and baryon chemical potential $\mu_B$.

\begin{figure}
    \centering
    \includegraphics[width=\columnwidth]{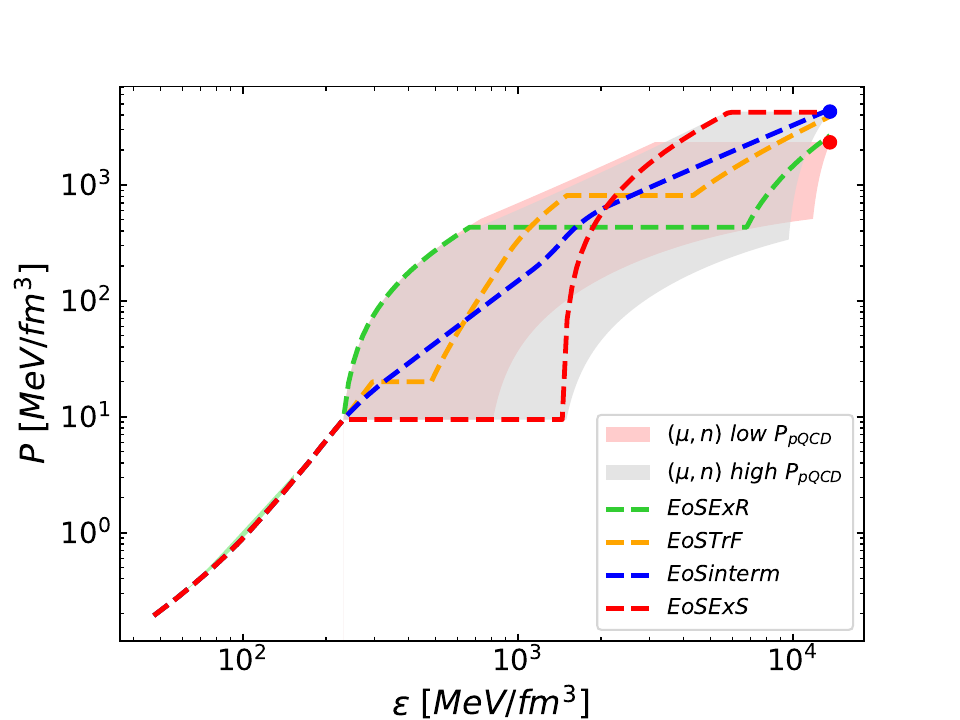}
    \caption{[Color online] Example set of four EoSs with $\chi$EFT data from \cite{Drischler:2020yad} matched to the intermediate region at $n= 1.5n_s$ and pQCD limits from \cite{Gorda:2021znl}. Two uncertainty bands appear, consequence of projecting a tube in several dimensions.  The foreground (grey) and background (red) areas correspond to simultaneously imposing the limit for $n$ and $\mu_B$  at either the high-$\varepsilon$ pQCD or the low-$\varepsilon$ pQCD  points, respectively. 
 The green line is the stiffest EoS in the neutron star density region, while the red line represents the softest allowed EoS there.} 
    \label{fig:EoSinterpol}
\end{figure}

 In this study, we present results from the interpolation at $1.5n_s$. From the band to be sampled we have selected four representative EoSs: the stiffest and softest band limits, and two intermediate ones, as shown in Fig.~\ref{fig:EoSinterpol}.
 
We denote by  EoSExS  the stiffest one at low densities in the interpolated region.  It is extracted by applying the maximum slope $c_s^2 \approx 1$ from the matching point between the interpolation and the chiral regions (at $n=1.5n_s$) to the point where the chemical potential is approximately 2100 MeV ($\varepsilon=667$ MeV/fm$^3$, $P=432$ MeV/fm$^3$). At this point, a long phase transition is applied until $\varepsilon=6773$ MeV/fm$^3$ and from here until entering pQCD by using $c_s^2 \approx 1/3$. We cannot construct a stiffer EoS with the known constraints.

EoSExS is in turn the softest EoS (referring to low densities in the interpolated region) and it is constructed by applying the minimum slope $c_s^2= 0$ from $n=1.5n_s$ to $n=9.44 n_s$ ($\varepsilon= 1446$ MeV/fm$^3$, $P=9.465$ MeV/fm$^3$). From this point, the maximum slope $c_s^2 \approx 1$ is applied until we achieve a chemical potential $\mu \approx$ 2600 MeV ($\varepsilon=5764$ MeV/fm$^3$, $P=4164$ MeV/fm$^3$). At this point a phase transition is applied to match into the pQCD band (entering it at $\varepsilon=13600$ MeV/fm$^3$, $P=4164$ MeV/fm$^3$).  

The two intermediate EoS have been constructed by very smoothly increasing the slope from  the point $n=1.5 n_s$. In the first case (EoSTrF), a first long phase transition has been used from $\varepsilon=295$ MeV/fm$^3$ ( $P=20$ MeV/fm$^3$) to $\varepsilon=483$ MeV/fm$^3$. From this point, we increase the slope to reach the maximum value ($c_s^2 \approx 1$) and then mantain this maximum slope until ($\varepsilon=1505$ MeV/fm$^3$, $P=811$ MeV/fm$^3$) where a second phase transition to $\varepsilon=4329$ MeV/fm$^3$ is applied. From the end of this phase transition, the conformal slope $c_s^2 \approx 1/3$ is used until the pQCD regime has been reached.

In the second intermediate EoS (EoSinterm), the same slope has been used as in the matching point with chiral region $n=1.5 n_s$ until $n=2.37 n_s$. The slope is then increased to $c_s^2= 0.62$ ($\varepsilon=1506$ MeV/fm$^3$, $P=365$ MeV/fm$^3$), decreased again to $c_s^2 \approx 1/3$  and maintained until pQCD is reached. These choices of the intermediate behaviour have no known microscopic meaning and are just representative of plausible EoS that satisfy all microscopic constraints.

\begin{figure}
    \centering
    \includegraphics[width=\columnwidth]{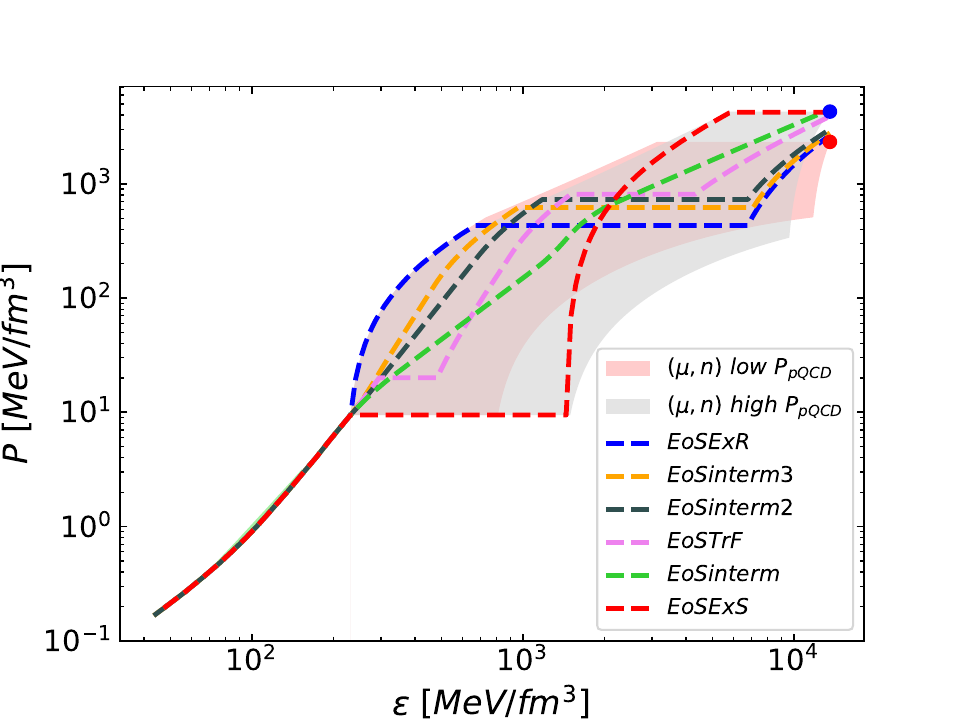}
    \caption{[Color online] Similar to figure~\ref{fig:EoSinterpol} but with additional intermediate 
    equations of state.} 
    \label{fig_6EoS}
\end{figure}

We then add an additional pair of possible such EoS in figure~\ref{fig_6EoS} that make the plot a bit busier.

These are the precise equations of state that are used throughout the investigation of rotating stars, for example
in figure~\ref{fig:RingofStatic} that shows results corresponding to precisely these six equations of state.

\end{document}